\documentclass[preprint, 3p]{elsarticle}

\usepackage{lipsum}
\makeatletter
\def\ps@pprintTitle{%
 \let\@oddhead\@empty
 \let\@evenhead\@empty
 \def\@oddfoot{}%
 \let\@evenfoot\@oddfoot}
\makeatother

\usepackage{amssymb}
\usepackage{url}
\usepackage{amsthm}
\usepackage{amsbsy}
\usepackage{amsmath}
\usepackage{algorithm,algpseudocode}

\usepackage{xcolor}
\usepackage{makecell}
\usepackage{tabularx}

\usepackage{lineno,hyperref}
\modulolinenumbers[5]

\begin{document}

\begin{frontmatter}

\title{Real-time simulation of parameter-dependent fluid flows \\ through deep learning-based reduced order models}

\author[1]{Stefania Fresca}
\ead{stefania.fresca@polimi.com}
\author[1]{Andrea Manzoni\corref{cor1}}
\ead{andrea1.manzoni@polimi.com}
\cortext[cor1]{Corresponding author}

\address[1]{MOX - Dipartimento di Matematica, Politecnico di Milano, P.zza Leonardo da Vinci 32, 20133 Milano, Italy}

\begin{abstract}
Simulating fluid flows in different virtual scenarios is of key importance in engineering applications. However, high-fidelity, full-order models relying, e.g., on the finite element method, are unaffordable whenever fluid flows must be simulated in almost real-time. Reduced order models (ROMs) relying, e.g., on proper orthogonal decomposition (POD)  provide reliable approximations to parameter-dependent fluid dynamics problems in rapid times. However, they might require expensive hyper-reduction strategies for handling parameterized nonlinear terms, and enriched reduced spaces (or Petrov-Galerkin projections) if a mixed velocity-pressure formulation is considered, possibly hampering the evaluation of reliable solutions in  real-time. Dealing with fluid-structure interactions entails even higher difficulties. The proposed deep learning (DL)-based ROMs overcome  all these limitations by learning in a non-intrusive way both the nonlinear trial manifold and the reduced dynamics. To do so, they rely on deep neural networks, after performing a former dimensionality reduction through POD  enhancing  their training times substantially. %, and a multi-fidelity pretraining stage. 
The resulting POD-DL-ROMs are shown to provide accurate results in almost real-time for the flow around a cylinder benchmark, the fluid-structure interaction between an elastic beam attached to a fixed, rigid block and a laminar incompressible flow, and the blood flow in a cerebral aneurysm.
\end{abstract}

\begin{keyword}
fluid dynamics; deep learning; reduced order modeling; proper orthogonal decomposition; Navier-Stokes equations; fluid-structure interaction
\end{keyword}

\end{frontmatter}
 
%%%%%%%%%%%%%%%%%%%%%%%%%%%%%%%%%%%%%%%%%%

\section{Introduction}

Computational fluid dynamics nowadays provide rigorous and reliable tools for the numerical approximation of fluid flows equations, exploited in several fields, from life sciences to aeronautical engineering. High-fidelity techniques such as, e.g.,  finite elements, finite volumes as well as spectral methods have been extensively applied in the past decades to the simulation of challenging problems in fluid dynamics, providing quantitative indication about the physical behavior of the system, in view of its better understanding, control, and forecasting. Solving these problems entails the numerical approximation of unsteady Navier-Stokes (NS) equations in three-dimensional domains, possibly accounting for fluid-structure interaction (FSI) effects; this requires  fine computational meshes in case one aims at simulating complex flow patterns, and ultimately yielding large-scale nonlinear systems of equations to be solved.  
Simulating fluid flows in complex configurations through high-fidelity, full-order models (FOMs) is computationally infeasible if one aims at solving the problem multiple times, for different virtual scenarios, or in a very small amount of time -- at the limit, in real-time. This is the case, for instance, of blood flow simulations, for which outputs of clinical interest shall be evaluated for different flow conditions, and in different geometrical configurations \cite{quarteronimanzonivergara2017}. In this respect, if quantitative outputs are meant to support clinicians' decisions, each new numerical simulation should be carried out very rapidly on deployed platforms, rather than exploiting huge parallel hardware architectures, and thus requiring limited data storage and memory capacity. 

In the case virtual scenarios can be described in terms of -- e.g., physical and/or geometrical -- input parameters, reduced order models (ROMs) built, e.g., through the reduced basis (RB) method \cite{quarteroni2015reduced}, can be exploited to reduce the computational complexity and costs entailed by the repeated solution of parametrized fluid flow problems, enabling dramatic reduction of the dimension of the discrete problems, arising from numerical approximation, from millions to hundreds, or thousands at most, of variables. Several works have addressed the construction of rapid and reliable ROMs for Navier-Stokes equations, mainly exploiting either proper orthogonal decomposition (POD) \cite{kunisch2002galerkin,GunzburgerPetersonShadid,BERGMANN2009516,weller2010numerical,ballarin2015supremizer,DM18}  or greedy algorithms \cite{VeroyPatera,deparis2008reduced,manzoni2014efficient,yano2014space} for the construction of reduced order spaces. However, despite the general principles behind projection-based reduction techniques -- such as, e.g., the use of a (Petrov-)Galerkin projection onto a low-dimensional subspace, and the use of a set of FOM snapshots, computed for different input parameter values, at different times, to train the ROM -- that provide a rigorous framework to set up ROMs for fluid dynamics equations, some distinguishing properties of Navier-Stokes equations for incompressible flow simulations ultimately make their effective realization quite involved \cite{lassila2014model}. Among them, we mention the need of {\em (i)} treating efficiently nonlinearities and parameter dependencies \cite{carlberg2013gnat}, {\em (i)} approximating both velocity and pressure \cite{caiazzo2014numerical}, {\em (iii)} ensuring the ROM stability (with respect to both  the violation of the inf-sup condition and dominating convection) \cite{carlberg2011efficient,BCI13}, and {\em (iv)} keeping error propagation in time under control. The presence of FSI, coupling the fluid model with a model describing the structural displacement of the non-rigid domain where the fluid flows, makes the problem even more involved. 
Several strategies have been proposed to address these issues: for instance, hyper-reduction techniques have been devised in a purely algebraic way to treat the nonaffine and nonlinear convective terms appearing in the NS equations \cite{DM18}; a suitable enrichment of the velocity space can be considered to ensure the inf-sup stability of the ROM \cite{rozza2013reduced,ballarin2015supremizer}, as well as alternative, more effective, stabilization techniques for the ROM \cite{dalsanto2017rbStokes}; mesh-moving techniques have been exploited to efficiently parametrize domain shapes to address geometric variability in fluid flow simulations \cite{DM18}, and either monolithic or segregated strategies have been considered as first attempts to handle fluid-structure interactions in the RB method for parametrized fluid flows \cite{lassila2013reduced,colciago2014comparisons}.  %

On the other hand, machine learning techniques -- in particular, artificial neural networks (NNs) -- in computational fluid dynamics have witnessed a dramatic blooming in the past ten years \cite{brunton2020machine,kutz2017deep}. Deep neural networks (DNNs) have been exploited to address several issues; a nonexhaustive list includes, for instance:
\begin{enumerate}

\item the extraction of relevant flow features, such as recirculation regions or boundary layers %and horseshoe vortices in a 3D wing-body junction flow 
through convolutional neural networks (CNNs) \cite{Strofer_2019};

\item the construction of  inexpensive, non-intrusive approximations for output quantities of interest for fluid flows \cite{lye2020deep}, or to velocity and pressure field, obtained through  Reynolds-averaged Navier-Stokes (RANS) equations \cite{bhatnagar2019prediction,thuerey2020deepFlowPred,Eichinger2020};

\item data-driven turbulence models in RANS equations through a physics-informed machine learning approach %for reconstructing discrepancies in RANS-modeled Reynolds stresses 
\cite{wang2019prediction}, or data-driven eddy viscosity closure models in Large Eddy Simulations (LES) \cite{beck2019deep};

\item the setting of closure models to stabilize a POD-Galerkin ROM \cite{san2018neural} by using, e.g., recurrent neural networks (RNNs) to predict the impact of the unresolved scales on the resolved scales \cite{wang2020recurrent}, or correction models to adapt a ROM to describe scenarios quite far from the ones seen during the training stage \cite{baiges2020finite};

\item the reconstruction of a high-resolution flow field from limited  flow information \cite{Raissi_Science}, as well as the assimilation of flow measurements and computational flow dynamics models derived from first physical principles. This task can be cast in the framework of the so-called physics-informed neural networks \cite{raissi2019physics,kissas2020machine}, where NNs are trained to solve supervised learning tasks while respecting the fluid dynamics equations, or tackled by means of Bayesian neural networks \cite{sun2020physics};

\item the nonintrusive estimation of POD coefficients through, e.g., feedforward NNs \cite{hestaven2018non-intrusive,wang2019non,san2019artificial} or probabilistic NNs \cite{fukami2020probabilistic}.

\end{enumerate}

In this paper we apply the POD-DL-ROM technique we recently proposed \cite{fresca2020POD}  to fluid flow problems, in order to build non-intrusive and extremely efficient ROMs for parameter-dependent unsteady problems in computational fluid dynamics by exploiting {\em (i)} deep neural networks as main building block, {\em (ii)} a set of FOM snapshots, and {\em (iii)} dimensionality reduction of FOM snapshots through (randomized) POD. Despite a preliminary example of its application to a benchmark in fluid dynamics has already been considered to assess the capability of POD-DL-ROMs to handle vector nonlinear problems such as the unsteady Navier-Stokes equations, in order to compute the fluid velocity field only, in this paper we deepen our analysis by considering: {\em (a)} the computation of both velocity and pressure fields in the case of unsteady Navier-Stokes equations, 
{\em (b)} the extension to a FSI problem, and 
{\em (c)} the application to a real-life application of interest, namely the simulation of blood flows through a cerebral aneurysm.  \\

Compared to other  works appeared recently in the literature, our focus is on parameter-dependent fluid dynamics problems, either involving complex three-di\-men\-sio\-nal geometries, or FSI effects, and on the use of deep learning (DL)-based ROMs for the sake of real-time simulation of fluid flows, thus relying on nonlinear reduction techniques. Motivated by similar goals, non-intrusive ROMs for fluid dynamics equations have been proposed, e.g., in \cite{xiao2015non,xiao2015non2,xiao2016non}, where POD has been considered to generate low-dimensional (linear) subspaces, also in the case of FSI problems, and POD coefficients at each time-step are either  computed through a radial basis function multi-dimensional interpolation, or  extrapolated from the POD coefficients at earlier time-steps. 

%rudy2019deep
%
%
%prediction of high-dimensional complex dynamical systems using neural networks and their variants
%
%Spectral proper orthogonal decomposition is applied to reduce the dimensionality of the model and, at the same time, filter the proper orthogonal decomposition temporal modes. The regression step is performed by a deep feedforward neural network (DNN)
%
%similar to the SINDy algorithm
%The framework is implemented in a context similar to that of the sparse identification of nonlinear dynamics (SINDy) algorithm (Brunton et al. 2016).

Applications of DL algorithms in conjunction with POD have already been proposed  for the sake of long-term predictions in time, however without addressing parameter-dependent problems. For instance, a long-short term memory (LSTM) network was used to learn the underlying physical dynamics in \cite{wang2018model}, generating a non-intrusive ROM through the solution snapshots acquired over time. Deep feedforward neural networks (DFNNs) have been used for a similar task in \cite{lui_wolf_2019} and compared with the sparse identification of nonlinear dynamics (SINDy) algorithm \cite{Brunton3932}. This latter defines a sparse representation through a linear combination of selected functions, and has been used for data-driven forecasting in fluid dynamics \cite{rudy2019deep}. RNNs have been considered in \cite{gonzalez2018deep,bukka2021assessment} to evolve low-dimensional states of unsteady  flows, exploiting either POD or a convolutional recurrent autoencoder to extract low-dimensional features from snapshots. DL algorithms have also been used to describe the reduced trial manifold where the approximation is sought, then relying on a minimum residual formulation to derive the ROM -- hence, still requiring the assembling and the solution of a ROM as in traditional POD-Galerkin ROMs -- in \cite{carlberg2018model}.  \\

The structure of the paper is as follows. In section \ref{sec:methods} we sketch the basic features of projection-based ROMs for fluid flows, and recall the main ingredients of the POD-DL-ROM technique. In Section \ref{sec:results} we show some numerical results obtained for the flow around a cylinder benchmark, the fluid-structure interaction between an elastic beam attached to a fixed, rigid block and a laminar incompressible flow, and the blood flow in a cerebral aneurysm. Finally, a brief discussion of our results and few comments about future research directions are reported in Section \ref{sec:discussion}.

% The introduction should briefly place the study in a broad context and highlight why it is important. It should define the purpose of the work and its significance. The current state of the research field should be reviewed carefully and key publications cited. Please highlight controversial and diverging hypotheses when necessary. Finally, briefly mention the main aim of the work and highlight the principal conclusions. As far as possible, please keep the introduction comprehensible to scientists outside your particular field of research. Citing a journal paper \cite{ref-journal}. Now citing a book reference \cite{ref-book1,ref-book2} or other reference types \cite{ref-unpublish,ref-communication,ref-proceeding}. Please use the command \citep{ref-thesis,ref-url} for the following MDPI journals, which use author--date citation: Administrative Sciences, Arts, Econometrics, Economies, Genealogy, Histories, Humanities, IJFS, Journal of Intelligence, Journalism and Media, JRFM, Languages, Laws, Religions, Risks, Social Sciences.
 
%%%%%%%%%%%%%%%%%%%%%%%%%%%%%%%%%%%%%%%%%%
\section{Methods}\label{sec:methods}

In this section we briefly recall the main ingredients of the POD-enhanced DL-based ROMs (briefly, POD-DL-ROMs) that we adapt, in the following, to handle problems in computational fluid dynamics. In particular, we aim at simulating parameter-dependent unsteady fluid flows, relying on a velocity-pressure formulation, in domains that have either {\em (i)} rigid walls or {\em (ii)} elastic deformable walls. 

In the case of rigid walls, for any input parameter vector $\boldsymbol{\mu} \in \mathcal{P} \subset \mathbb{R}^{n_{\boldsymbol{\mu}}}$, we aim at solving the nonlinear unsteady Navier-Stokes equations in a given, fixed domain $\Omega^F \subset \mathbb{R}^d$, $d=2,3$ (see Section \ref{sec_31})
\begin{equation}
%\left\{
%\begin{array}{lll}
P_F({\bf v}_h , p_h  ; t, \boldsymbol{\mu}) = 0   \qquad  \mbox{in} \ \Omega^F \times (0,T), 
%\mbox{boundary conditions} & \qquad & on \  \partial \Omega^F \times (0,T), \\
%\mbox{initial condition} & \qquad & on \  \Omega^F \times \{ t = 0 \} .
%\end{array}
%\right.
\label{eq:fluid_problem}
\end{equation}
in the time interval $(0,T)$,  provided that suitable initial (at time $t=0$) and boundary  conditions (on $\partial \Omega^F$, for each $t \in (0,T)$) are assigned.
Here $t \in (0,T)$ is the time variable, ${\bf v}_h = {\bf v}_h (t; \boldsymbol{\mu})$ the velocity field, $p_h = p_h (t; \boldsymbol{\mu})$ the pressure field; these two latter quantities are usually obtained through  a FOM built, e.g., through the finite element method. Here $h>0$ denotes a discretization parameter, usually related to the mesh size.

In the case of elastic walls, the fluid domain is unknown and its deformation introduces a further geometric nonlinearity;  the structure displacement ${\bf d}_h^S = {\bf d}_h^S (t; \boldsymbol{\mu})$ might also be nonlinear, and must match the one of the fluid domain ${\bf d}_h^G = {\bf d}_h^G (t; \boldsymbol{\mu})$ at the fluid-structure (FS) interface $\Sigma(t)$. Here we employ the so-called Arbitrary Lagrangian Eulerian (ALE) approach, in which an extra problem for the fluid domain displacement  (usually a harmonic extension of the FS interface datum) is solved, thus providing an updated fluid domain, while the fluid problem is reformulated on a frame of reference that moves with the fluid domain. Thus, for any input parameter vector $\boldsymbol{\mu} \in \mathcal{P}$, we consider a  fluid-structure interaction (FSI) model, which consists of a two-fields problem, coupling the incompressible Navier-Stokes equations written in the ALE  form with the (non)linear elastodynamics equation modeling the solid deformation \cite{fsi_wiley}. %Because of the employed ALE approach, a third mesh motion problem is introduced, accounting for the fluid domain deformation, and defining the ALE map.
 In particular, we aim at solving the unsteady Navier-Stokes equations in a varying domain $\Omega^F(t)  \subset \mathbb{R}^d$, the elastodynamics equations in the structural domain $\Omega^S \subset \mathbb{R}^d$, and a geometric problem in the fixed fluid domain $\Omega^F  \subset \mathbb{R}^d$ (see Section \ref{sec_32}),  %provided suitable initial (at time $t=0$) and boundary  conditions on $\partial \Omega^F$:
\begin{equation}
\left\{
\begin{aligned}
P_{F}({\bf v}_h , p_h  ; t, \boldsymbol{\mu}) = 0 & \qquad & \mbox{in} \  \Omega^F(t)  \times (0,T),\\
P_S ( {\bf d}_h^S; t, \boldsymbol{\mu}) = 0 & \qquad &  \mbox{in} \  \Omega^S  \times (0,T), \\
P_G ( {\bf d}_h^G ;   \boldsymbol{\mu}) = 0 & \qquad &  \mbox{in} \  \Omega^F  \times (0,T), \\
\mbox{coupling conditions} & \qquad &  \mbox{on} \ \Sigma(t) \times (0,T), \\
%\mbox{boundary conditions} & \qquad & on \  \partial \Omega^F(t) \setminus \Sigma \times (0,T) \\
%\mbox{initial conditions} & \qquad & on \  \Omega^F \cup \Omega^S \times \{ t = 0 \}   
\end{aligned}
\right.
\label{eq:fsi_problem}
\end{equation}
for a time interval $(0,T)$, provided that suitable initial (at time $t=0$) and boundary  conditions (on $\partial \Omega^F(t) \setminus \Sigma(t)$ for the fluid subproblem, on $\partial \Omega^S \setminus \Sigma$ for the structural subproblem, on $\partial \Omega^F$ for the geometric subproblem,   for each $t \in (0,T)$), are assigned.

\subsection{Projection-based ROMs: main features}

The spatial discretization of problem \eqref{eq:fluid_problem} or \eqref{eq:fsi_problem} through finite elements yields a nonlinear dynamical system of dimension $N_h$ to be solved for each input parameter value; then, a fully discretized problem is obtained relying, e.g., on either semi-implicit or implicit methods introducing a partition of the interval $[0, T ]$ in $N_t$ subintervals of equal size $\Delta t = T /N_t$, such that $t^k = k \Delta t$. This results in a sequence of either linear or nonlinear algebraic systems to be solved at each time step $t^k$, $k=1,\ldots,N_t$ -- which we refer to as the high-fidelity FOM.  
Note that the dimension $N_h$ accounts for the degrees of freedom of either the fluid problem (involving velocity and pressure) or the FSI problem (also including the structural and the geometrical subproblem). 
Building a projection-based ROM through, e.g., the RB method, then requires to perform this calculation for $n_s$ selected parameter values $\boldsymbol{\mu}^1, \ldots, \boldsymbol{\mu}^{n_s}$, and to perform POD on the solution snapshots (obtained for each $\boldsymbol{\mu}^j, j=1,\ldots, n_s$, and for each time step $t^k$, $k=1,\ldots,N_t$). Focusing for the sake of simplicity on the fluid problem \eqref{eq:fluid_problem}, the RB approximation of velocity and pressure fields at time $t^k$  is expressed as a linear combination of the RB basis functions,
\[
{\bf v}_h(t^k; \boldsymbol{\mu})  \approx  {\bf V}^v {\bf v}_N(t^k; \boldsymbol{\mu}), \qquad
{\bf p}_h(t^k; \boldsymbol{\mu})  \approx  {\bf V}^p {\bf p}_N(t^k; \boldsymbol{\mu})
\]
where ${\bf V}^v \in \mathbb{R}^{N_h \times N_v}$ and ${\bf V}^p \in \mathbb{R}^{N_h \times N_p}$ denote the matrices whose columns form the basis for the velocity and the pressure RB spaces, respectively, and are selected as the first left singular vectors of the (velocity and pressure) snapshots matrices. Note that in this case the RB approximation is sought in a linear trial manifold. A similar approximation also holds for the additional variables appearing in the FSI problem \eqref{eq:fsi_problem}. The reduced dynamics is then obtained by solving a low-dimensional dynamical system, obtained by performing a Galerkin projecting of the FOM onto the spaces spanned by the RB spaces; alternatively, a Petrov-Galerkin projection could also be used.  

Projection-based ROMs for parametrized PDEs thus rely on a suitable offline-online computational splitting: computationally expensive tasks required to build the low-dimensional subspaces, and to assemble the ROM arrays, are performed once for all in the so-called offline (or ROM training) stage. This latter allows us to compute -- ideally -- in an extremely efficient way the ROM approximation for any new parameter value, during the so-called online (or ROM testing) stage. This splitting, however, might be compromised if {\em (i)} the dimension of the linear trial subspace becomes very large (compared to the intrinsic dimension of the solution manifold being approximated), such as in the case of problems featuring coherent structures that propagate over time like transport, wave, or convection-dominated phenomena, or {\em (ii)}  hyper-reduction techniques, required to approximate $\boldsymbol{\mu}$-dependent nonlinear terms, require linear subspaces whose dimension is also very large. Even more importantly, two additional issues make the construction of ROMs quite critical in the case of fluid dynamics problems; indeed,
\begin{enumerate}
\item 
a Galerkin projection onto the RB space built through the POD procedure above does not  ensure the stability of the resulting ROM (in the sense of the fulfillment of an inf-sup condition at the reduced level). Several strategies can be employed to overcome this issue such as, e.g., {\em (a)} the augmentation of the velocity space by means of a set of enriching basis functions computed through the so-called pressure supremizing operator, which depends on the divergence term; {\em (b)} the use of a Petrov-Galerkin (e.g., least squares, (LS)) RB method, or {\em (c)} the use of a stabilized FOM (like, e.g., a P1-P1 Streamline Upwind Petrov-Galerkin (SUPG) finite element method); {\em (d)} an independent treatment of the pressure, to be reconstructed from the velocity by solving a Poisson equation, in the case divergence-free velocity basis functions are used -- an assumption that might be hard to fulfill;

\item the need of dealing with both a mixed formulation and a coupled FSI problem requires the construction of a reduced space for each variable, no matter if one is interested in the evaluation of output quantities of interest only involving a single variable. For instance, even if one is  interested in the evaluation of the fluid velocity in the FSI case, a projection-based ROM must account for all the variables appearing as unknowns in the coupled FSI problem. The same consideration also holds in the case of a fluid problem, where the pressure must be treated as unknown of the ROM problem even if one is not interested in its evaluation.
\end{enumerate}

\subsection{POD-enhanced DL-ROMs (POD-DL-ROMs)}

POD-DL-ROMs are nonintrusive ROMs, which aim at approximating the map $(t, \boldsymbol{\mu}) \rightarrow {\bf u}_h(t, \boldsymbol{\mu})$, for any field variable of interest ${\bf u}_h(t, \boldsymbol{\mu})$, by describing both the trial manifold and the reduced dynamics through deep neural networks. These latter are trained on a set of FOM snapshots
\begin{equation}\label{eq:snap_matrix1}
{\bf S}_{\bf u} = [   \mathbf{u}_h(t^1; \boldsymbol \mu_1) \; | \; \ldots \; | \; \mathbf{u}_h(t^{N_t}; \boldsymbol \mu_1) \; | \; \ldots \; | \; \ldots
\mathbf{u}_h(t^1 ; \boldsymbol \mu_{N_{train}}) \; | \; \ldots \; | \; \mathbf{u}_h(t^{N_t} ; \boldsymbol \mu_{N_{train}}) ],
\end{equation}
computed for different parameter values  $\boldsymbol{\mu}_1, \ldots, \boldsymbol{\mu}_{N_{train}} \in \mathcal{P}$, suitably sampled over the parameter space, at different time instants $\{t^1, \ldots, t^{N_t}\} \subset [0,T]$. Avoiding the {\em projection} stage, POD-DL-ROMs can be cheaply evaluated once trained, only involving those variables one is interested in. In case multiple variables are involved (e.g., both velocity and pressure), the procedure below can be performed simultaneously on each of them.

To reduce the dimensionality of the snapshots and avoid to feed training data of very large dimension $N_h$, we first apply POD -- realized through randomized SVD (rSVD) -- to the snapshot set ${\bf S}_{\bf u}$; then, a DL-ROM is built to approximate the map between $(t, \boldsymbol{\mu})$ and the POD generalized coordinates. 
Using rSVD, we build  $N$-dimensional subspace $\textnormal{Col}(\mathbf{V}_N)$ spanned by the $N \leq N_h$ columns of ${\bf V}_{N} \in \mathbb{R}^{N_h \times N}$, the matrix of the first $N$ singular vectors of the snapshot matrix ${\bf S}_u$. Here $N$ denotes  the dimension of the linear manifold, which can be taken (much) larger than the one of the reduced linear trial manifold used in a POD-Galerkin ROM. 

Hence,  the POD-DL-ROM approximation of the FOM solution ${\mathbf{u}}_h(t; \boldsymbol{\mu})$ is  
\begin{equation*}
\mathbf{\tilde{u}}_h(t; \boldsymbol{\mu}, \boldsymbol{\theta}_{DF}, \boldsymbol{\theta}_{D}) = \mathbf{V}_N \mathbf{\tilde{u}}_N(t; \boldsymbol{\mu}, \boldsymbol{\theta}_{DF}, \boldsymbol{\theta}_{D}) \approx {\mathbf{u}}_h(t; \boldsymbol{\mu}),
\end{equation*}
that is, it is sought in a linear trial manifold of (potentially large) dimension $N$, 
\begin{equation}
\label{manifold_linear_N}
\tilde{\mathcal{S}}_h^{N} = \{
{\bf V}_N \mathbf{\tilde{u}}_N(t; \boldsymbol{\mu}, \boldsymbol{\theta}_{DF}, \boldsymbol{\theta}_{D})  \ | \mathbf{\tilde{u}}_N(t; \boldsymbol{\mu}, \boldsymbol{\theta}_{DF}, \boldsymbol{\theta}_{D}) \in \mathbb{R}^N,    \ t \in [0, T)  \; , \boldsymbol{\mu} \in \mathcal{P}   \} \subset \mathbb{R}^{N_h},
\end{equation}
by applying the DL-ROM strategy \cite{fresca2020comprehensive} to approximate $\mathbf{V}_N^T\mathbf{u}_{h}(t ; \boldsymbol{\mu} )$ -- rather than directly ${\mathbf{u}}_h(t; \boldsymbol{\mu})$.  The DL-ROM approximation 
$\mathbf{\tilde{u}}_N(t; \boldsymbol{\mu}, \boldsymbol{\theta}_{DF}, \boldsymbol{\theta}_{D}) \approx \mathbf{V}_N^T\mathbf{u}_{h}(t ; \boldsymbol{\mu} )$ takes the form
\begin{equation}
\mathbf{\tilde{u}}_N(t; \boldsymbol \mu, \boldsymbol{\theta}_{DF}, \boldsymbol{\theta}_D ) = {\mathbf{f}}^D_N({\boldsymbol{\phi}}_n^{DF}(t; \boldsymbol{\mu}, {{\boldsymbol{\theta}}_{DF}}); \boldsymbol{\theta}_{D}),
\label{eq:u_N_approx}
\end{equation}
and is sought in a reduced nonlinear trial manifold $\tilde{\mathcal{S}}_N^n$ of very small dimension $n \ll N$; usually, $n \approx n_{\boldsymbol{\mu}} + 1$ -- here time is considered as an additional parameter. In particular:
\begin{itemize}
\item to describe the system dynamics on  the nonlinear trial manifold $\tilde{\mathcal{S}}_N^n$,  the intrinsic coordinates of the approximation $\mathbf{\tilde{u}}_N$ are defined as  
\begin{equation*}
\label{eq:phi_n}
\mathbf{u}_n(t; \boldsymbol \mu) = \boldsymbol{\phi}_n^{DF}(t; \boldsymbol \mu, \boldsymbol{\theta}_{DF}),
\end{equation*}
where   $\boldsymbol{\phi}_n(\cdot ;  \cdot, \boldsymbol{\theta}_{DF}) : [0, T) \times \mathbb{R}^{n_{\mu} +1} \rightarrow \mathbb{R}^n$ is a DFNN, 
consisting of the repeated composition of a nonlinear activation function,  applied to a linear transformation of the input, multiple times.  Here $\boldsymbol{\theta}_{DF}$ denotes the  DFNN {parameters} vector, collecting the weights and biases of each of its layers;
\item to model the  reduced nonlinear trial manifold $\tilde{\mathcal{S}}_N^n$, we  employ the decoder function of a convolutional autoencoder (CAE), that is, 
\begin{equation}
\begin{split}
\tilde{\mathcal{S}}_N^n = \{ \mathbf{\tilde{u}}_{N}(t; \boldsymbol \mu) = & {\mathbf{f}}^D_N({\boldsymbol{\phi}}_n^{DF}(t; \boldsymbol{\mu}, {{\boldsymbol{\theta}}_{DF}}); \boldsymbol{\theta}_{D}) \; | \\
& \; \mathbf{u}_n(t; \boldsymbol{\mu},  {{\boldsymbol{\theta}}_{DF}}) \in \mathbb{R}^{n}, \ t \in [0, T) \; , \; \boldsymbol{\mu} \in \mathcal{P} \subset \mathbb{R}^{n_{\mu}} \} 	\subset \mathbb{R}^N,
\end{split}
\label{eq:manifold_tilde_S_n}
\end{equation}
where ${\mathbf{f}}^D_N( \cdot ; \boldsymbol{\theta}_{D}) : 	\mathbb{R}^n \rightarrow \mathbb{R}^N$ denotes the   decoder function of a CAE obtained as   the composition of several layers (some of which are convolutional), depending upon a vector ${\boldsymbol{\theta}}_{D}$ collecting all the corresponding weights and biases.
\end{itemize}
Finally, the encoder function ${\mathbf{f}}^E_n( \cdot ; \boldsymbol{\theta}_{E}) : 	\mathbb{R}^N \rightarrow \mathbb{R}^n$  --  depending upon a vector ${\boldsymbol{\theta}}_E$ of parameters -- of the CAE can be used to map the intrinsic coordinates $\mathbf{V}_N^T \mathbf{u}_h (t, \boldsymbol{\mu})$ associated to $(t, \boldsymbol{\mu})$ onto a low-dimensional representation
\begin{equation*}
\label{eq:f_n}
 {\mathbf{\tilde{u}}_n}(t; \boldsymbol{\mu}, \boldsymbol{\theta}_{E}) = {\mathbf{f}}_{n}^E(\mathbf{V}_N^T \mathbf{u}_h(t; \boldsymbol{\mu}); \boldsymbol{\theta}_{E}).
\end{equation*}
Hence, training a POD-DL-ROM requires to solve the optimization problem 	 
 \begin{equation}
\min_{\boldsymbol{\theta}} \mathcal{J}(\boldsymbol{\theta}) = \min_{\boldsymbol{\theta}} \frac{1}{N_{train} N_t}\sum_{i=1}^{N_{train}} \sum_{k=1}^{N_t} \mathcal{L}(t^k, \boldsymbol \mu_i; \boldsymbol{\theta}),  \label{eq:minimization_problem}
\end{equation}
where the \emph{per-example} loss function $\mathcal{L}(t^k, \boldsymbol{\mu}_i;  {\boldsymbol{\theta}})$  is given by the sum of two terms, 
\begin{equation}
\label{eq:loss_N}
\mathcal{L}(t^k, \boldsymbol{\mu}_i;  {\boldsymbol{\theta}})  = \frac{\omega_h}{2} \mathcal{L}_{rec}(t^k, \boldsymbol{\mu}_i;  {\boldsymbol{\theta}})  + \frac{1-\omega_h}{2}   
\mathcal{L}_{int}(t^k, \boldsymbol{\mu}_i;  {\boldsymbol{\theta}}); 
\end{equation}
the former is the reconstruction error  between the FOM  and the POD-DL-ROM solutions,  
\[
\mathcal{L}_{rec}(t^k, \boldsymbol{\mu}_i;  {\boldsymbol{\theta}}) = \| \mathbf{V}_N^T\mathbf{u}_h(t^k; \boldsymbol{\mu}_i) - \mathbf{\tilde{u}}_N(t^k; \boldsymbol{\mu}_i,  {\boldsymbol{\theta}_{DF}, \boldsymbol{\theta}_D})\|^2;
\]
the latter is the misfit between the  {intrinsic coordinates} and the output of the encoder, 
\[
\mathcal{L}_{int}(t^k, \boldsymbol{\mu}_i;  {\boldsymbol{\theta}}) =  \| \tilde{\mathbf{u}}_n(t^k; \boldsymbol{\mu}_i, \boldsymbol{\theta}_E) -  {\mathbf{u}}_n(t^k; \boldsymbol{\mu}_i,  {\boldsymbol{\theta}_{DF}})\|^2.  
\]
Finally, $\omega_h \in [0,1]$ is a prescribed weighting parameter. 

 Computing the POD-DL-ROM approximation $\mathbf{\tilde{u}}_h(t; \boldsymbol{\mu}_{test})$ of ${\mathbf{u}}_h(t; \boldsymbol{\mu}_{test})$, for any $t \in (0,T)$ and $\boldsymbol{\mu}_{test} \in \mathcal{P}$, corresponds to the testing  stage of the DFNN and of the decoder function of the CAE, and does not require the evaluation of the encoder function. Finally, the POD-DL-ROM approximation of the FOM solution  is  recovered  as 
\begin{equation*}
\mathbf{\tilde{u}}_h(t; \boldsymbol{\mu}, \boldsymbol{\theta}_{DF}, \boldsymbol{\theta}_{D}) = \mathbf{V} \mathbf{\tilde{u}}_N(t; \boldsymbol{\mu}, \boldsymbol{\theta}_{DF}, \boldsymbol{\theta}_{D}).
\end{equation*}

Let us remark that the former construction of a POD-DL-ROM can be extended to the case of $p > 1$ field variables of interest, in a straightforward way.  In this case, provided a snapshots set  ${\bf S}_i$, and a corresponding basis $\mathbf{V}_{N,i} \in \mathbb{R}^{N_{h,i} \times N}$, $i=1,\ldots, p$, for each of the variables ${\bf u}_{h,1}$, \ldots, ${\bf u}_{h,p}$, the POD-DL-ROM approximation of the field variable ${\bf u}_{h,i}(t; \boldsymbol{\mu}) \in \mathbb{R}^{N_{h,i}}$ is given by
\[
\mathbf{\tilde{u}}_{h,i}(t; \boldsymbol{\mu}, \boldsymbol{\theta}_{DF}, \boldsymbol{\theta}_{D}) = \mathbf{V}_{N,i} \mathbf{\tilde{u}}_{N,i}(t; \boldsymbol{\mu}, \boldsymbol{\theta}_{DF}, \boldsymbol{\theta}_{D}) \approx {\mathbf{u}}_{h,i}(t; \boldsymbol{\mu}),
\]
where a DFNN and a CAE {are trained by considering simultaneously all the} $p$ field variables. Due to its {\em data-driven} nature, each variable can be approximated in an independent way -- in other words, there are no physical constraints appearing in the loss function, thus making the $p$ approximated field variables uncoupled, despite they might be originally coupled.  For instance, $p=d+1$ variables are considered if we aim at approximating both the velocity and the pressure fields in the case of fluid flows. 

Another noteworthy aspect deals with the way snapshots are handled when considering convolutional layers in the NNs, in presence of either vector and/or coupled problems. Exploiting the analogy with red-green-black images in image processing, each snapshot computed for a variable of interest is reshaped in a square matrix of dimension $(\sqrt{N}, \sqrt{N})$, where $N = 2^{(2m)}$ with $m \in \mathbb{N}$ (if $N \neq 2^{(2m)}$ the input is
zero-padded), and stacked together forming a tensor with $p$ {\em channels}. The latter tensor is then provided as input to the POD-DL-ROM neural network architectures  when dealing with vector and/or coupled problems; as a result, the output of the network, for each sample $(t, \boldsymbol{\mu})$, takes a form similar to (\ref{eq:u_N_approx}), collecting the approximation of all the field variables,
\[
\mathbf{\tilde{u}}_N (t; \boldsymbol \mu, \boldsymbol{\theta}_{DF}, \boldsymbol{\theta}_D ) = [\mathbf{\tilde{u}}_{N,1}(t; \boldsymbol \mu, \boldsymbol{\theta}_{DF}, \boldsymbol{\theta}_D )
 \; | \; \ldots \; | \; 
\mathbf{\tilde{u}}_{N,p}(t; \boldsymbol \mu, \boldsymbol{\theta}_{DF}, \boldsymbol{\theta}_D )
 ]  \in \mathbb{R}^{N \times p}.
\]
%\textcolor{red}{
%\begin{itemize}
%\item riempimento con zeri fino ad arrivare a $N = 2^{(2m)}$?
%\item altre cose da aggiungere?
%\end{itemize}
%}
We remark that considering vector and/or coupled problems does not entail main changes in the architecture of the POD-DL-ROM, as well as in the total number of parameters of the neural network. Indeed, only the first layer of the encoder function and the last one of the decoder function are responsible for the handling of the different channels of the input/output. This implies that training the neural network by providing data with $p$ channels is remarkably less computationally expensive than training $p$ independent POD-DL-ROMs, each of them responsible for a single component/field variable of the solution.

\section{Results}\label{sec:results}

In this section, we show several numerical results obtained with the POD-DL-ROM technique. In particular, we focus on the solution of three problems: (\textit{i}) the unsteady Navier-Stokes equations for a two-dimensional  flow around a cylinder,  (\textit{ii}) a FSI problem for a two-dimensional flow past an elastic beam attached to a fixed, rigid block and (\textit{ii}) the unsteady Navier-Stokes equations for the blood flow in a cerebral aneurysm. To evaluate the performance of POD-DL-ROM, we rely on the loss function (\ref{eq:loss_N}) and on:
\begin{itemize}
\item the error indicator $\epsilon_{rel} \in \mathbb{R}$ given by
\begin{equation}
\epsilon_{rel} = \epsilon_{rel}(\mathbf{u}_h, \mathbf{\tilde{u}}_h) = \frac{1}{N_{test}} \sum_{i  = 1}^{N_{test}} \left(\displaystyle \frac{\sqrt{ \sum_{k=1}^{N_t} || \mathbf{u}^k_h(\boldsymbol{\mu}_{test,i}) - \mathbf{\tilde{u}}^k_h(\boldsymbol{\mu}_{test,i}) ||^2}}{\sqrt{\sum_{k=1}^{N_t} || \mathbf{u}_h^k(\boldsymbol{\mu}_{test,i}) ||^2}} \right);
\label{eq:error_indicator}
\end{equation}
\item the relative error $\boldsymbol{\epsilon}_k \in \mathbb{R}^{\sum_{i=1}^d N_{h,i}}$, for $k = 1, \ldots, N_t$, defined as
\begin{equation}
\displaystyle \boldsymbol{\epsilon}_k = \displaystyle \boldsymbol{\epsilon}_k(\mathbf{u}_h, \mathbf{\tilde{u}}_h) = \displaystyle \frac{ | \mathbf{u}^k_h(\boldsymbol{\mu}_{test}) - \mathbf{\tilde{u}}^k_h(\boldsymbol{\mu}_{test}) |}{\sqrt{\frac{1}{N_t}\sum_{k=1}^{N_t} || \mathbf{u}^k_h(\boldsymbol{\mu}_{test}) ||^2}}.
\label{eq:relative_error_s}
\end{equation}
\end{itemize}
Note that  (\ref{eq:error_indicator}) is a scalar indicator, while (\ref{eq:relative_error_s}) provides  a spatially-distributed error field.

The configuration of the POD DL-ROM neural network
used in our test cases is the one given below. We choose a 12-layer DFNN equipped with 50 neurons per hidden layer and $n$ neurons in the output layer, where $n$ represents the dimension of the (nonlinear) reduced  {trial} manifold. The architectures of the encoder and decoder functions are instead  reported in \tablename s \ref{table_transposed_convolutional_layers_encoder} and \ref{table_transposed_convolutional_layers}.  No activation function is applied at the last convolutional layer of the decoder neural network, as usually done when dealing with AEs. 
\begin{table}[h!]
\caption{Attributes of convolutional  and dense layers in the encoder $\mathbf{f}_n^E$.}
\label{table_transposed_convolutional_layers_encoder}
{\small
\begin{center}
\begin{tabular}{|c|c|c|c|c|c|c|}
\hline
layer & input & output  & kernel  & $\#$of filters & stride & padding \\
  & dimension &   dimension &   size &  &  &   \\
\hline
1 & [N, N, $d$] & [N, N, 8] & [5, 5] & 8 & 1 & SAME \\
\hline
2 & [N, N, 8] & [N/2, N/2, 16] & [5, 5] & 16 & 2 & SAME \\
\hline
3 & [N/2, N/2, 16] & [N/4, N4, 32] & [5, 5] & 32 & 2 & SAME \\
\hline
4 & [N/4, N/4, 32] & [N/8, N/8, 64] &[5, 5] & 64 & 2 & SAME \\
\hline
5 & N & 64 & & & & \\
\hline
6 & 64 & $n$ & & & &\\
\hline
\end{tabular}
\end{center}
}
\vspace{-0.5cm}
\end{table}
\begin{table}[h!]
\caption{Attributes of dense  and transposed convolutional layers in the decoder $\mathbf{f}_N^D$.}
\label{table_transposed_convolutional_layers}
{\small
\begin{center}
\begin{tabular}{|c|c|c|c|c|c|c|}
\hline
layer & input & output  & kernel  & $\#$of filters & stride & padding \\
  & dimension &   dimension &   size &  &  &   \\
  \hline
1 & $n$ & 256 & & & &\\
\hline
2 & 256 & $N_h$ & & & &\\
\hline
3 & [N/8, N/8, 64] & [N/4, N/4, 64] & [5, 5] & 64 & 2 & SAME \\
\hline
4 & [N/4, N/4, 64] & [N/2, N/2, 32] & [5, 5] & 32 & 2 & SAME \\
\hline
5 & [N/2, N/2, 32] & [N, N, 16] & [5, 5] & 16 & 2 & SAME \\
\hline
6 & [N, N, 16] & [N, N, $d$] & [5, 5] & $d$ & 1 & SAME \\
\hline
\end{tabular}
\end{center}
}
\end{table} 

To solve the optimization problem \eqref{eq:minimization_problem}-\eqref{eq:loss_N}, we use the ADAM algorithm \cite{kingma2015adam}, which is a stochastic gradient descent method computing an adaptive approximation of the first and second momentum of the gradients of the loss function. In particular, it computes exponentially weighted moving averages of the gradients and of the squared gradients. We set the starting learning rate to $\eta = 10^{-4}$, and  perform cross-validation in order to tune the hyperparameters of the POD-DL-ROM, by splitting the data in training and validation sets with a proportion 8:2. Moreover, we implement an early-stopping  regularization technique  to reduce overfitting \cite{goodfellow2016deep}, stopping the training if the loss does not decrease over a certain amount of epochs. As nonlinear activation function, we employ the ELU function \cite{clevert2015fast}. The parameters, weights and biases, are initialized through the He uniform initialization \cite{he2015delving}. The interested reader can refer to \cite{fresca2020POD} for a more in-depth description of these architectures, and for a detailed  version of the algorithms used for the training and the testing phases. These latter have been carried out on a Tesla V100 32GB GPU for the cases described in the following subsections.  The \texttt{Matlab} library \texttt{redbKIT} \cite{quarteroni2015reduced,redbKIT} has been employed to carry out all the FOM simulations.

\subsection{Test case 1: flow around a cylinder}\label{sec_31}

\begin{figure}[t!]
\centering
%\vspace{-0.25cm}
\includegraphics[scale=0.375]{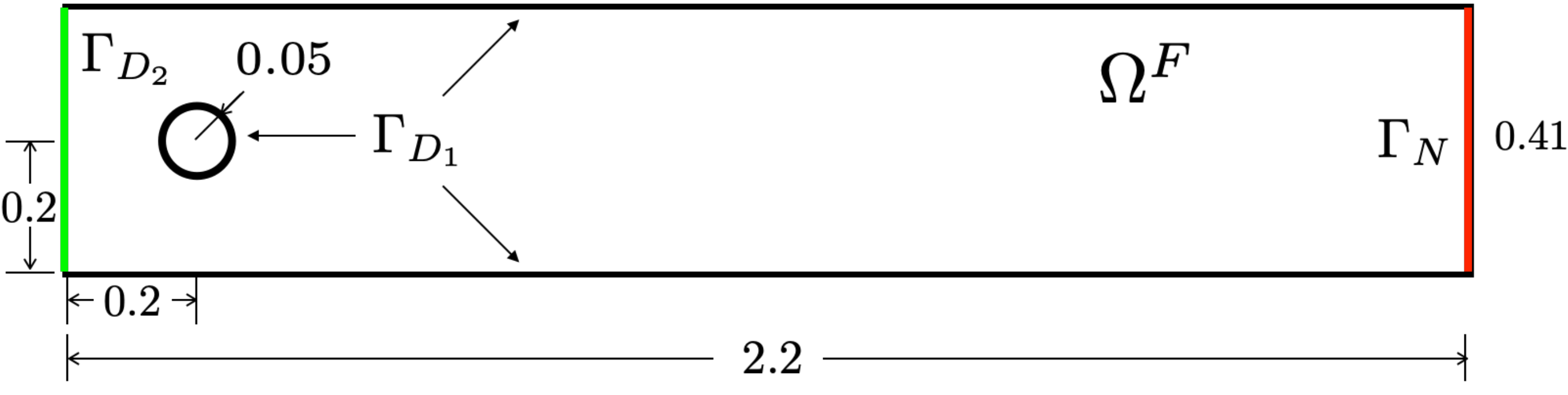}
\caption{\textit{Test case 1}: geometrical configuration, domain and boundaries  {[m]}.}
\label{fig:Fig5-11}
\end{figure}

In this first test case we deal with the unsteady Navier-Stokes equations for incompressible flows in primitive variables (fluid velocity $\mathbf{v}$ and pressure $p$). We consider the flow around a cylinder test case, a well-known benchmark problem for the evaluation of numerical algorithms for incompressible Navier-Stokes equations in the laminar case \cite{schafer1996benchmark}. The problem reads as follows:
\begin{equation}
\left\{
\begin{aligned}
& \rho \frac{\partial \mathbf{v}}{\partial t} + \rho \mathbf{v} \cdot \nabla \mathbf{v}   - \nabla \cdot \boldsymbol{\sigma}(\mathbf{v}, p) =  \mathbf{0} & \qquad  & (\mathbf{x},t) \in \Omega^F \times (0,T), \\
&\nabla \cdot \mathbf{v} = 0  & \qquad & (\mathbf{x},t) \in \Omega^F \times (0,T),\\
& \mathbf{v} =  \mathbf{0} & \qquad & (\mathbf{x},t) \in \Gamma_{D_1} \times (0,T),\\
& \mathbf{v} =  \mathbf{h} & \qquad & (\mathbf{x},t) \in \Gamma_{D_2} \times (0,T),\\
& \boldsymbol{\sigma}(\mathbf{v}, p) \mathbf{n}= \mathbf{0} & \qquad & (\mathbf{x},t) \in \Gamma_{N}\times (0,T), \\
&  \mathbf{v}(0) = \mathbf{0} & \qquad & \mathbf{x} \in \Omega^F, \ t = 0.
\end{aligned}
\right.
\label{eq:NS}
\end{equation}
The domain consists in a two-dimensional pipe with a circular obstacle, i.e. $\Omega^F = (0, 2.2) \times (0, 0.41) \char`\\ \bar{B}_{0.05}(0.2,0.2)$ -- here ${B}_{r}({\bf x}_c)$ denotes a ball of radius $r>0$ centered at ${\bf x}_c$, see Figure~\ref{fig:Fig5-11} for a sketch of the geometry. The boundary is given by $\partial \Omega^F = \Gamma_{D_1} \cup \Gamma_{D_2} \cup \Gamma_{N}$, where   
$\Gamma_{D_1} = \{ x_1 \in [0, 2.2], x_2 = 0\} \cup \{ x_1 \in [0, 2.2], x_2 = 0.41\} \cup \partial B_{0.05}( (0.2, 0.2) ) $,  $\Gamma_{D_2} = \{ x_1 = 0, x_2 \in [0, 0.41] \}$, and  $\Gamma_N = \{x_1 = 2.2, x_2 \in [0, 0.41] \}$, while $\mathbf{n}$ denotes the (outward directed) normal unit vector to $\partial \Omega^F$. We denote by 
 $\rho$ the fluid density, and by $\boldsymbol{\sigma}$ the stress tensor, 
\begin{equation}
\label{eq:sigma}
\boldsymbol{\sigma}(\mathbf{v}, p) = -p \mathbf{I} + 2 \nu \boldsymbol{\epsilon}(\mathbf{v}); 
\end{equation}
here $\nu$ denotes the dynamic viscosity of the fluid, while $\boldsymbol{\epsilon}(\mathbf{v})$ is the strain tensor, 
\begin{equation*}
\boldsymbol{\epsilon}(\mathbf{v}) = \frac{1}{2} \big( \nabla \mathbf{v} + \nabla \mathbf{v} ^T \big). 
\end{equation*}
% \cite{NS}).

Here we take  $\rho=1$ kg/m$^3$ as fluid density, and assign no-slip boundary conditions on $\Gamma_1$, a parabolic inflow profile  
\begin{equation}
\label{eq:h}
\mathbf{h}(\mathbf{x},t; \mu) = \left( \frac{4 U(t, \mu) x_2 (0.41-x_2)}{0.41^2} , 0 \right), \qquad \mbox{where } \ \  U(t; \mu) = \mu \sin(\pi t / 8),
\end{equation}
on  the inlet $\Gamma_{D_2}$, and zero-stress Neumann conditions on the outlet $\Gamma_N$.  We consider as parameter ($n_\mu=1$)  $\mu \in \mathcal{P}=[1,2]$ m/s, which reflects on the Reynolds number varying in the range $[66,133]$. Equations (\ref{eq:NS}) have been discretized in space by means of linear-quadratic $(\mathbb{P}_2-\mathbb{P}_1$), inf-sup stable, finite elements, and in time through a backward differentiation formula (BDF) of order 2 with semi-implicit treatment of the convective term (see, e.g., \cite{forti2015semi}) over the time interval $(0,T)$, with $T=8$ s, and a time-step $\Delta t = 2 \times 10^{-3}$ s. This strategy allows us to mitigate the computational cost associated to the use of a fully implicit BDF scheme, by linearizing the nonlinear convective terms; this latter task is realized by extrapolating the convective velocity through an extrapolation formula of the same order of the BDF introduced.

We already analyzed this test case in \cite{fresca2020POD}, where we were interested in approximating only the velocity field. Here we aim at assessing the performance of POD-DL-ROM neural network in approximating both the velocity and the pressure fields. In particular, we provide to the network data under the form of a tensor with 3 channels -- that is,  we set the dimension equal to $p = 3$. The FOM dimension is equal to $N_h = [32446, 32446, 8239]$ (for the two velocity components and the pressure, respectively) and we select $N = 256$ as  dimension of the POD basis  for each component of the solution. We choose $n = 5$ as  dimension of the nonlinear trial manifold $\tilde{S}_n$. We uniformly sample $N_t =400$ time instances and consider $N_{train} = 21$ and $N_{test} = 20$ training- and testing-parameter instances uniformly distributed over $\mathcal{P}$.

\begin{figure}[h!t]
\vspace{-0.25cm}
%\widefigure
\centerline{
\includegraphics[width=8.5 cm]{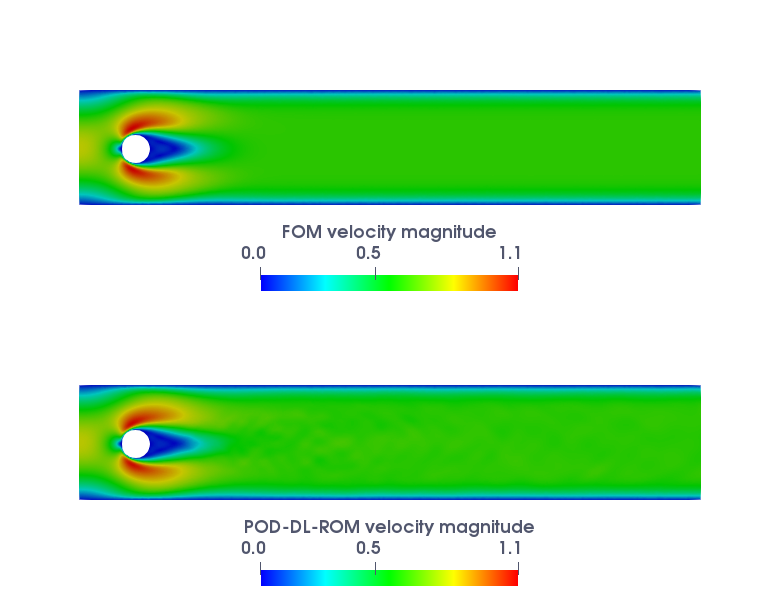} \hspace{-0.35cm}
\includegraphics[width=8.5 cm]{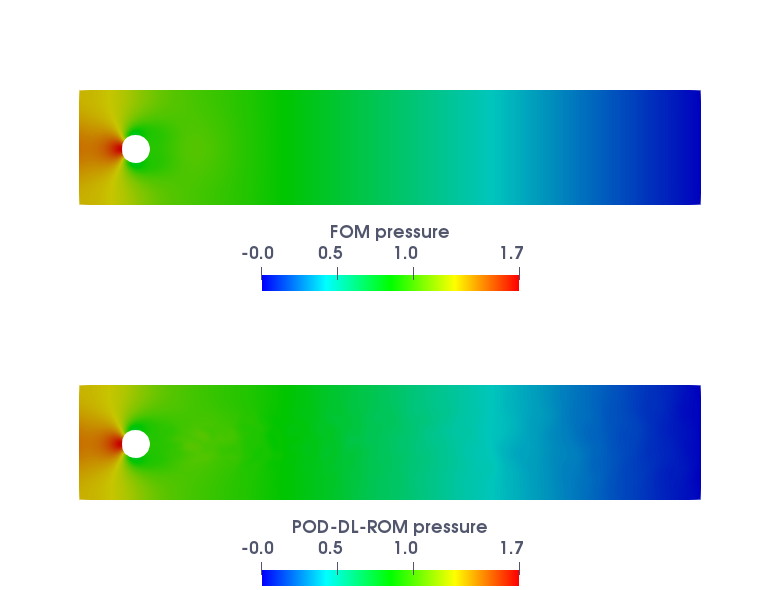}
}
\caption{\textit{Test case 1:} FOM and POD-DL-ROM solutions for the testing-parameter instance $\mu_{test} = 1.975$ m/s at $t =1.062 $ s, with $N = 256$ and $n = 5$. Left: velocity field magnitude; right: pressure field.}
\label{fig2}
\vspace{-0.35cm}
\end{figure}

\begin{figure}[h!t]
%\widefigure
\centerline{
\includegraphics[width=8.5 cm]{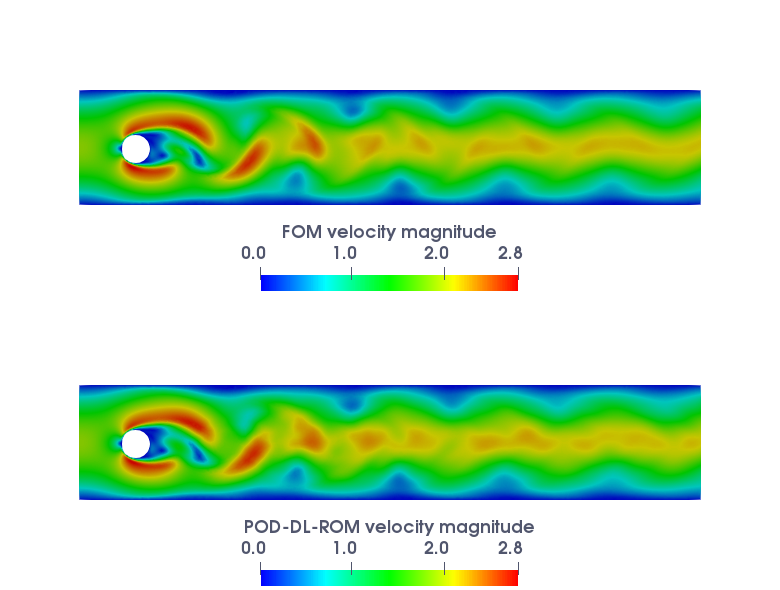} \hspace{-0.35cm}
\includegraphics[width=8.5 cm]{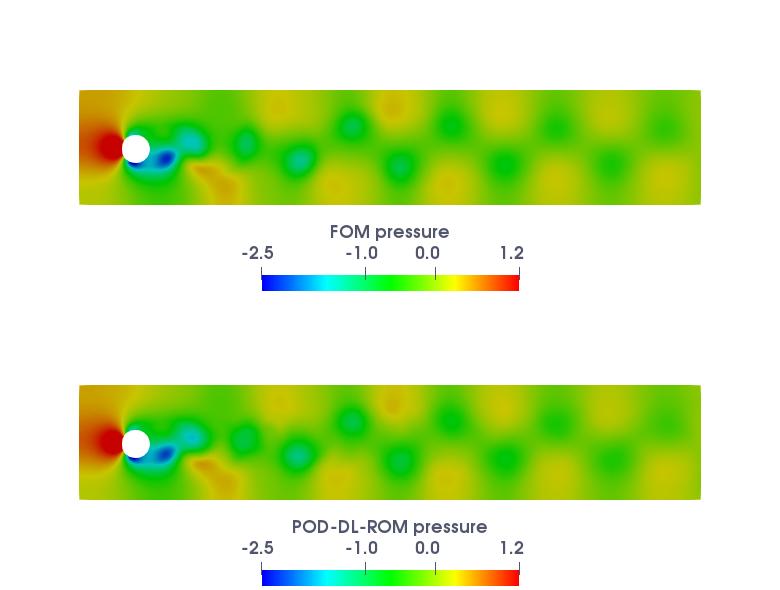}
}
\caption{\textit{Test case 1:} FOM and POD-DL-ROM solutions for the testing-parameter instance $\mu_{test} = 1.975$ m/s at $t = 4.842$ s, with $N = 256$ and $n = 5$. Left: velocity field magnitude; right: pressure field.}
\label{fig3}
\vspace{-0.25cm}
\end{figure}

\begin{figure}[ht]
%\widefigure
\centerline{
\includegraphics[width=8.5 cm]{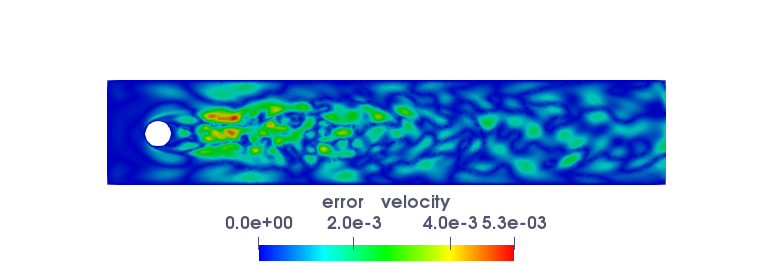} \hspace{-0.35cm}
\includegraphics[width=8.5 cm]{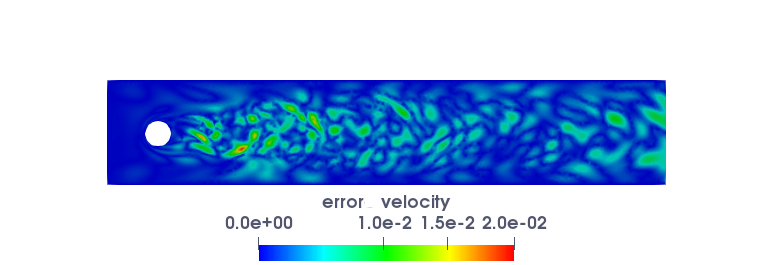}
}
\centerline{
\includegraphics[width=8.5 cm]{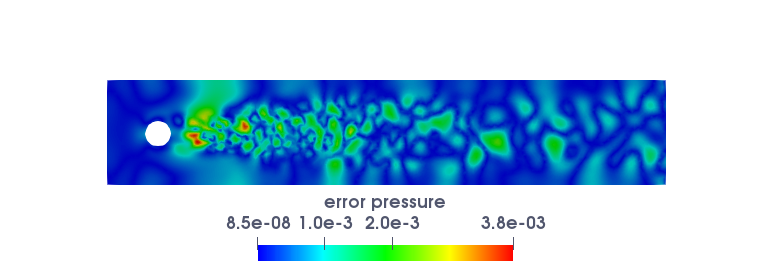} \hspace{-0.35cm}
\includegraphics[width=8.5 cm]{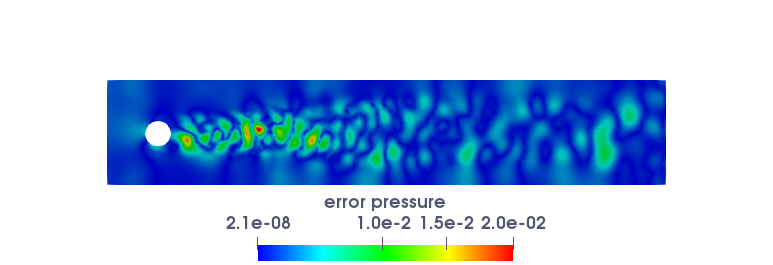}
}
\caption{\textit{Test case 1:} Relative errors $\epsilon_k$ for the testing-parameter instance $\mu_{test} = 1.975$ m/s, with $N = 256$ and $n = 5$. Relative errors at $t = 1.062$ s (left) and  $t = 4.842$ s (right) for velocity (left) and pressure (right).}
\label{fig4}
\end{figure}

In Figure~\ref{fig2}  and Figure~\ref{fig3}  we compare the FOM and POD-DL-ROM pressure and velocity fields, these latter obtained with $n = 5$, together with the relative error $\boldsymbol{\epsilon}_k$ in Figure~\ref{fig4}, for the testing-parameter instance $\mu_{test} = 1.975$ m/s ($Re = 131$) at $t = 1.062$ s and $t = 4.842$ s. We highlight the  ability of the POD-DL-ROM approximation to accurately capture the variability of the solution: indeed, in the case $t = 1.062$ s (Figure~\ref{fig2}) we do not assist to any vortex shedding; this latter is instead present in the case $t = 4.842$ s (Figure~\ref{fig3}), and is correctly reproduced. 

%\end{paracol}

%\begin{paracol}{2}
%\linenumbers
%\switchcolumn

To assess the ability of the POD-DL-ROM to provide accurate evaluations of output quantities of interest, we evaluate the drag and lift coefficients, related to the drag and lift forces around the circular obstacle; these are defined, in our case, as
\begin{equation}
\label{eq:forces}
F_D = \int_{\partial B_r} \Big( \nu \frac{\partial u_2}{\partial \eta} \eta_2 - p \eta_1 \Big) d\boldsymbol{\sigma}, \qquad \textnormal{and} \qquad F_L = \int_{\partial B_r} \Big( \nu \frac{\partial u_1}{\partial \eta} \eta_1 - p \eta_2 \Big) d\boldsymbol{\sigma} .
\end{equation}
where $\boldsymbol{\eta} = (\eta_1,\eta_2)^T$ denotes the (outward directed) normal unit vector to $\partial \Omega$. 
From (\ref{eq:forces}), the dimensionless drag and lift coefficient can be obtained as
\begin{equation*}
C_D = \frac{2}{U_{mean}^2L}F_D, \qquad \textnormal{and} \qquad C_L = \frac{2}{U_{mean}^2L}F_L,
\end{equation*}
where $U_{mean}$ is the parabolic input profile mean velocity.

\begin{figure}[ht]
%\widefigure
\includegraphics[width=8.25 cm]{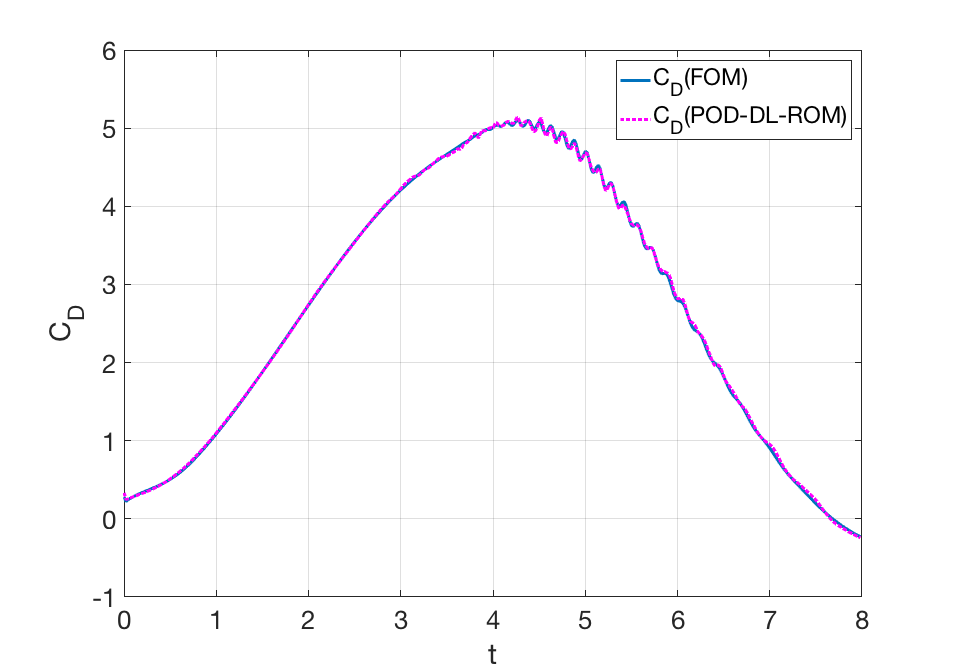}
\includegraphics[width=8.25 cm]{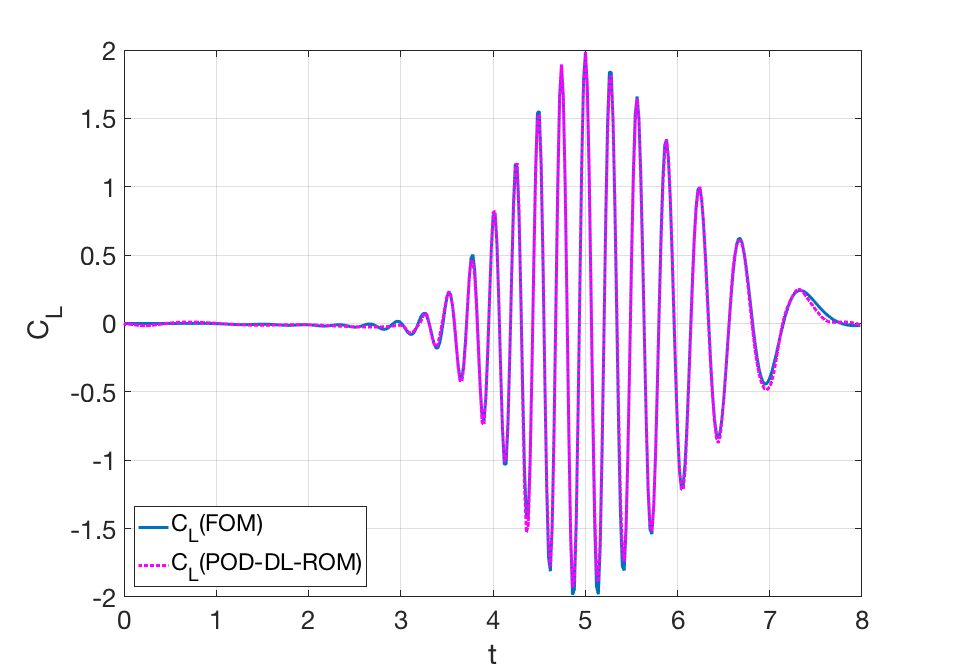}
\caption{\textit{Test case 1:} FOM and POD-DL-ROM drag (left) and lift (right) coefficients for the testing-parameter instance $\mu_{test} = 1.975$. }
\label{fig5}
\end{figure}

 \textcolor{black}{The drag and lift coefficients coefficients computer over time by the   FOM and the POD-DL-ROM, for the testing-parameter instances $\mu_{test} = 1.975$ m/s, are reported in Figure~\ref{fig5}. The POD-DL-ROM technique is also able to accurately capture the evolution of $C_D$ and $C_L$, related to the prescribed $\mu$-dependent input profile in (\ref{eq:h}), in both cases.} 

%\end{paracol}
%\begin{paracol}{2}
%\linenumbers
%\switchcolumn

Finally, the testing computational time, i.e. the time needed to compute $N_t$ time instances for an unseen testing-parameter instance, of the POD-DL-ROM on a GTX 1070 8GB GPU is given by 0.2 seconds, thus implying a speed-up equal to $1.25 \times 10^5$ with respect to the time needed for the solution of the FOM\footnote{For test case 1, the FOM simulations have been performed on 20 cores of 1.7 TB node (192 Intel\textsuperscript{\textregistered} Xeon Platinum\textsuperscript{\textregistered} 8160 2.1GHz cores) of the HPC cluster available at MOX, Politecnico di Milano.}. 

\subsection{Test case 2: fluid-structure interaction}\label{sec_32}

\begin{figure}[h!t]
\centering
\includegraphics[width=9.25 cm]{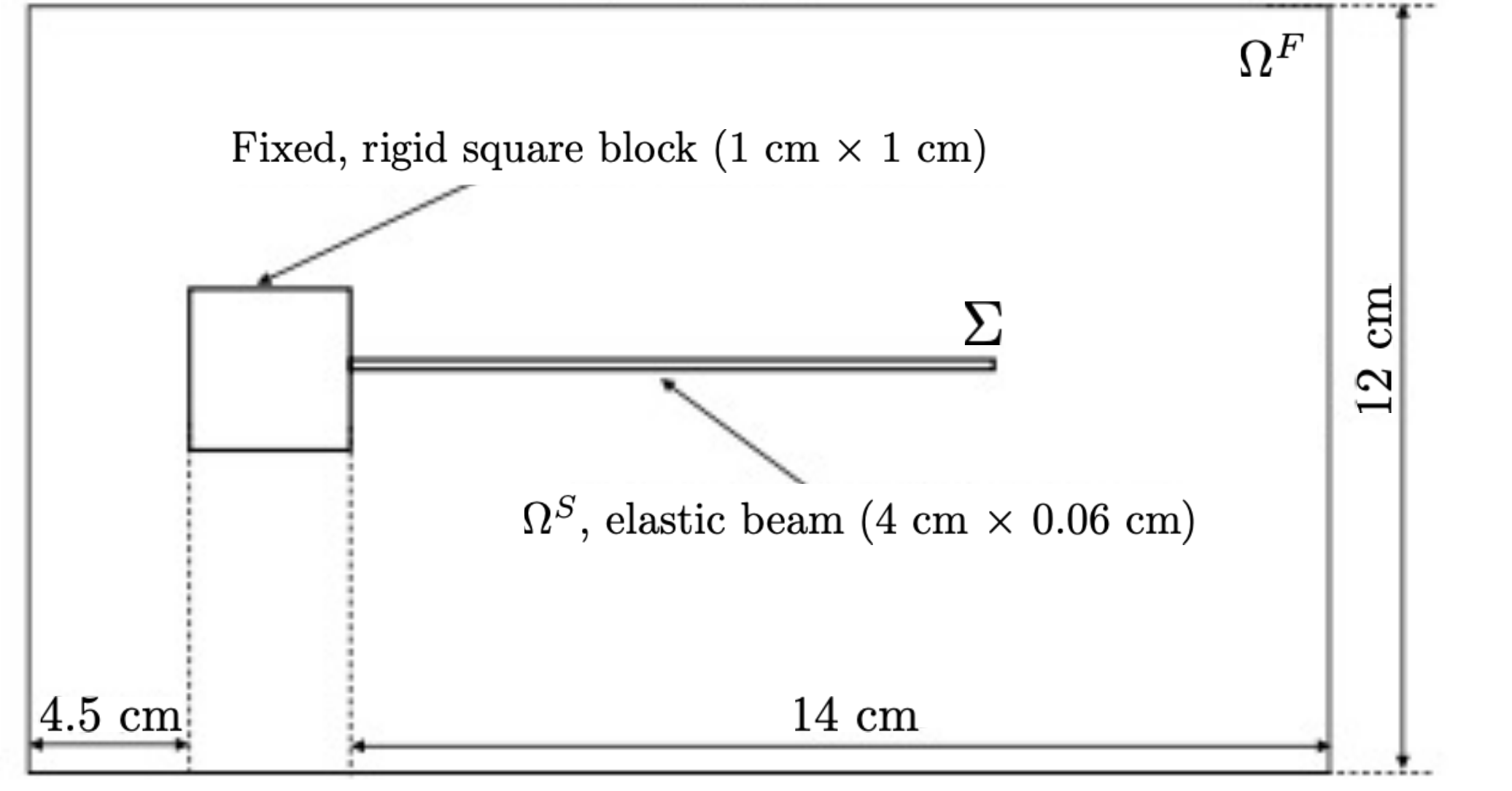} \\
\caption{\textit{Test case 2:} geometrical configuration and domains.}
\label{fig6}
\end{figure}

We now focus on the case of a two-dimensional flow past an elastic beam attached to a fixed, rigid block \cite{wall1999fluid, wall1998fluid, bazilevs2008isogeometric} (see Figure~\ref{fig6} for a sketch of the geometry). The FSI model that we consider consists of a two-fields problem, where the incompressible Navier-Stokes equations written in the Arbitrary Lagrangian Eulerian (ALE) form for the fluid are coupled with the nonlinear elastodynamics equation modeling the solid deformation \cite{fsi_wiley}. Because of the ALE approach we employ, a third non-physical geometry (or mesh motion) problem is introduced, which accounts for the fluid domain deformation and in turn defines the so-called ALE map.

Let $\Omega^F$ and $\Omega^S$ be the domains occupied by the fluid and the solid, respectively, in their reference configuration. We denote by $\textcolor{black}{\Sigma} = \partial \Omega^F \cap \partial \Omega^S$ the fluid-structure interface on the reference configuration. At any time $t$, the domain occupied by the fluid $\Omega^F(t)$ can be retrieved from $\Omega^F$ by the ALE mapping
\begin{equation*}
\mathcal{A}_t : \,\, \Omega^F \rightarrow \Omega^F(t),  \qquad
\mathbf{X} \mapsto \mathcal{A}_t(\mathbf{X}) =  \mathbf{X} + \mathbf{d}^G(\mathbf{X}),
\end{equation*}
where $\mathbf{d}^G(\mathbf{X})$ represents the displacement of the fluid domain. The coupled FSI problem thus consists in the following set of equations:
\begin{itemize}
\item Navier-Stokes in ALE form governing the fluid problem: \vspace{-0.1cm}
\begin{equation}
\begin{cases}
\displaystyle
\rho^F \frac{\partial\mathbf{v}^F}{\partial t} \Big|_{\mathbf{X}} + (\mathbf{v}^F -\mathbf{w}^G) \cdot \nabla \mathbf{v}^F - \nabla \boldsymbol{\sigma}^F (\mathbf{v}^F, p^F) = 0 & \quad \textnormal{in} \; \Omega^F(t), \\
\nabla \cdot \mathbf{v}^F = 0 & \quad \textnormal{in} \; \Omega^F(t);
\end{cases}
\end{equation}
\item nonlinear elastodynamics equation governing the solid dynamics: \vspace{-0.1cm}
\begin{equation}
\displaystyle
\rho^S \frac{\partial \mathbf{d}^S}{\partial t^2} - \nabla \cdot \mathbf{P}(\mathbf{d}^S) = 0 \qquad \textnormal{in} \; \Omega^S;
\end{equation}
\item coupling at the FS interface $\Sigma$: \vspace{-0.1cm}
\begin{equation}
\begin{cases}
\displaystyle
\mathbf{v}^F = \frac{\partial \mathbf{d}^S}{\partial t}, \\
\boldsymbol{\sigma}^F (\mathbf{v}^F, p^F) \mathbf{n}^F + \boldsymbol{\sigma}^S(\mathbf{d}^S)\mathbf{n}^S = {\bf 0};
\end{cases}
\end{equation}
\item linear elasticity equations modeling the mesh motion problem: \vspace{-0.1cm}
\begin{equation}
\begin{cases}
\displaystyle
-\nabla \cdot \boldsymbol{\sigma}^G(\mathbf{d}^G) = 0 & \quad \textnormal{in} \; \Omega^F, \\
\mathbf{d}^G = \mathbf{d}^S & \quad \textnormal{on} \; \Sigma,
\end{cases}
\end{equation}
\end{itemize}
where $\boldsymbol{\sigma}^F(\mathbf{v}^F, p^F) = -p^F \mathbf{I} + 2 \mu^F \boldsymbol{\epsilon}(\mathbf{v}^F)$ is the fluid Cauchy stress tensor, $\boldsymbol{\sigma}^S(\mathbf{d}^S) = J^{-1} \mathbf{P}\mathbf{F}^T$ is the solid Cauchy stress tensor, $\mathbf{F} = {\bf I} + \nabla {\bf d}^S$ is the deformation tensor, and 
\begin{equation*}
\displaystyle
\mathbf{w}^G = \frac{\partial \mathbf{d}^G}{\partial t} \Big|_{\mathbf{X}}
\end{equation*}
is the fluid mesh velocity. See, e.g., \cite{fsi_wiley} for further details about the FSI model. 

Both fluid and solid equations are complemented by appropriate initial and boundary conditions. In particular, the lateral boundaries are assigned zero normal velocity and zero tangential stress. Zero-traction boundary condition is applied at the outflow. The flow is driven by a uniform inflow velocity of 51.3 cm/s. Zero-initial conditions are assigned both for the fluid and the solid equations. The fluid density and viscosity are $1.18 \times 10^{-3}$ g/cm$^3$ and $1.82 \times 10^{-4}$ g/(cm $\cdot$ s) respectively, resulting in a Reynolds number of 100 based on the edge length of the block. The beam is modeled as a solid made of the neo-Hookean material and the density of the beam is 0.1 g/cm$^3$. 

The field equations are discretized in space and time using:  matching spatial discretizations between fluid and structure at the interface;   SUPG stabilized linear
finite elements ($(\mathbb{P}_1-\mathbb{P}_1$) and a BDF of order 2 in time for the fluid subproblem;  the same finite element space as for the fluid velocity for both the fluid displacement subproblem and the structural subproblem, and the  Newmark scheme in time for this latter. The resulting nonlinear problem is solved through a  monolithic geometry-convective explicit (GCE) scheme,  obtained by linearizing the fluid convective term (with a BDF extrapolation) and treating the geometry problem explicitly \cite{crosetto2011parallel,gee2011truly}. Here $n_\mu = 2$ parameters are considered,  the Young modulus $\mu_1$ and the Poisson ratio $\mu_2$, varying in the parameter space $\mathcal{P}= 10^{6} \cdot [2.3,2.7] \; \textnormal{g/(cm $\cdot$ s$^2$)} \times [0.3, 0.4]$. We build a FOM considering $N_h = [16452, 8226, 1974]$ DOFs for the velocity, the pressure and the displacement fields, respectively, and a time step  $\Delta t = 1.65 \times 10^{-3}$ over  $(0,T)$ with $T=3$ s. 

Regarding the construction of the proposed POD-DL-ROM, for the training of the neural networks, we uniformly sample   $N_t = 606$ time instances   and $N_{train} = 5 \times 3 = 15$ training-parameter instances, uniformly distributed in each parametric direction. At testing phase,  $N_{test} = 4 \times 2 = 8$ testing-parameter instances, different from the training ones, have been considered. The maximum number of epochs is set equal to $N_{epochs} =20000$, the batch size is $N_b = 40$ and, regarding the early-stopping criterion, we stop the training if the loss function does not decrease within 500 epochs.  We are interested in reconstructing the velocity and the displacement fields, so we set the number of channels to {$p=2d=4$} and we recall the ability of the POD-DL-ROM neural network to handle different FOM dimensions $N_{h,i}$, for $i= 1, \ldots, p$, that is only the POD dimension must be equal among the different fields considered. {Moreover, we set $N = 256$ as dimension of the POD basis, and $n =5$ as  dimension of the reduced nonlinear trial manifold.} The training and testing phases of the POD-DL-ROM neural network have been performed on a Tesla V100 32GB GPU. 

\begin{figure}[h!]
\centering
\vspace{-0.25cm}
\includegraphics[width=12.5cm]{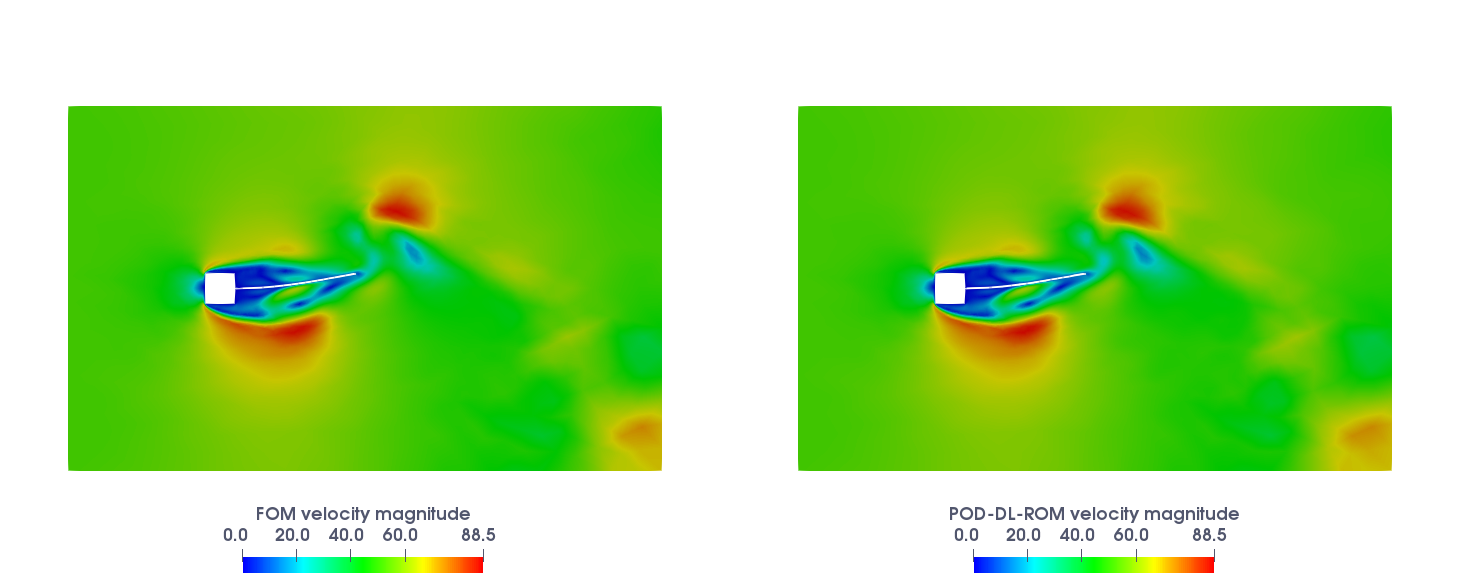} \\
\includegraphics[width=12.5cm]{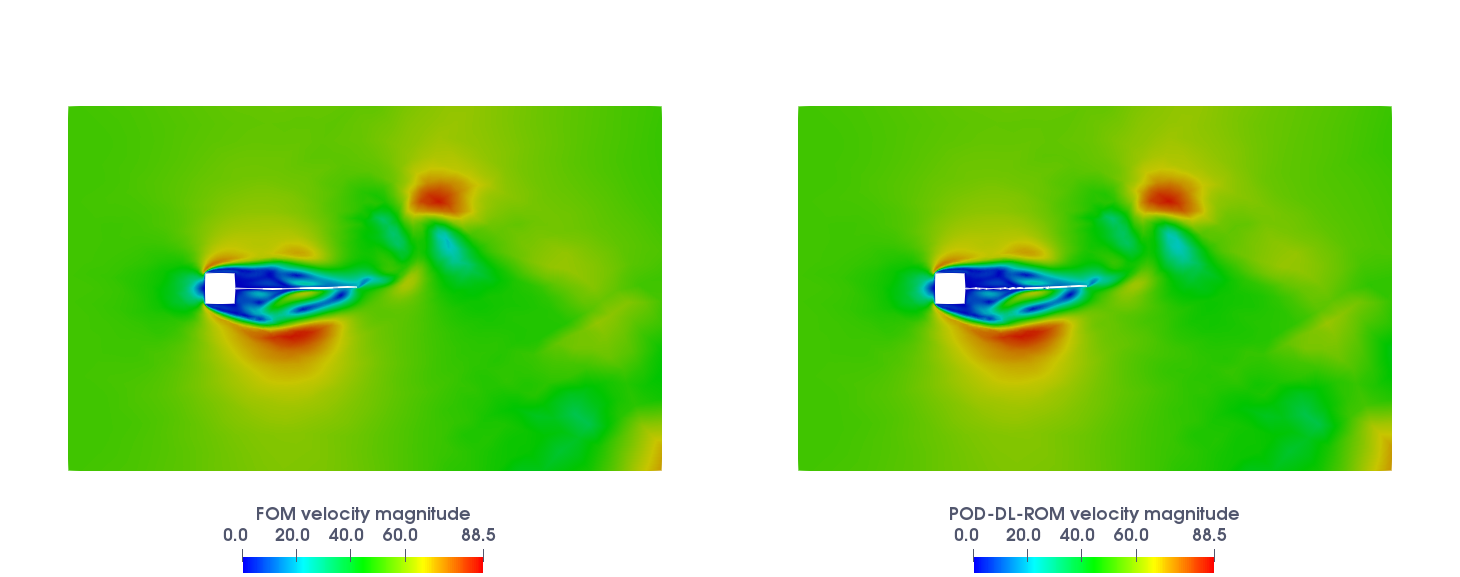} \\
\includegraphics[width=12.5cm]{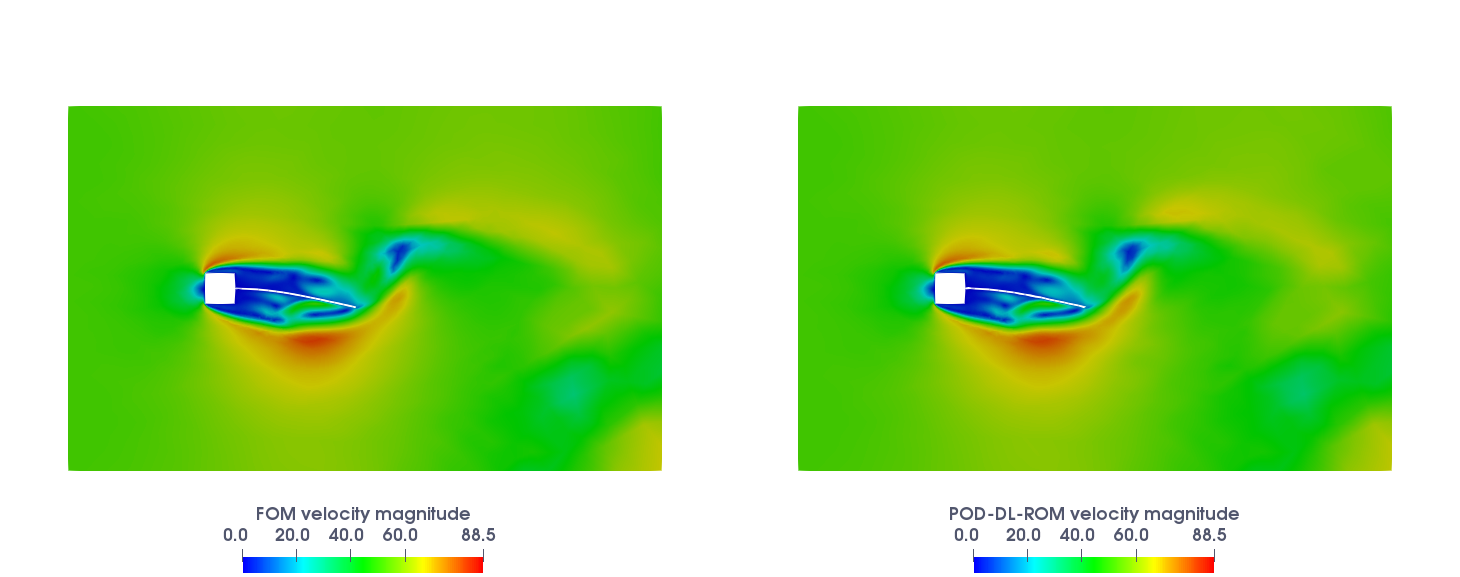} \\
\includegraphics[width=12.5cm]{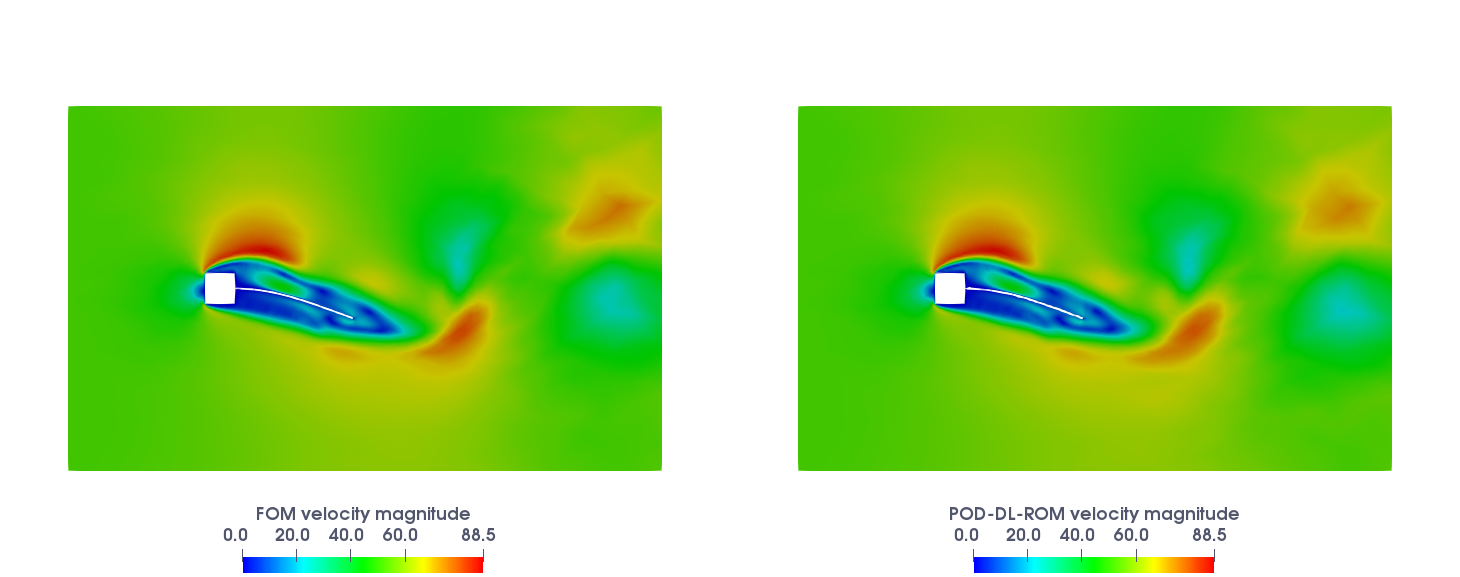} \\
\caption{\textit{Test case 2:} FOM and POD-DL-ROM fluid velocity magnitudes for the testing-parameter instances $\boldsymbol{\mu}_{test} = [2.3 \times 10^6 \textnormal{ g/(cm $\cdot$ s$^2$)}, 0.325]$ at $t = 2.3084$ s and $t = 2.64$ s (first and second row) and $\boldsymbol{\mu}_{test} = [2.7 \times 10^6 \textnormal{ g/(cm $\cdot$ s$^2$)}, 0.375]$ at $t = 2.3084$ s and $t = 2.64$ s (third and fourth row). }%, with $N = 256$ and $n = 5$. (\textbf{a}) $\boldsymbol{\mu}_{test} = [2.3 \times 10^6 \textnormal{ g/(cm $\cdot$ s$^2$)}, 0.325]$ and  $t = 2.3084$ s. (\textbf{b}) $\boldsymbol{\mu}_{test} = [2.3 \times 10^6 \textnormal{ g/(cm $\cdot$ s$^2$)}, 0.325]$ and  $t = 2.64$ s. (\textbf{c}) $\boldsymbol{\mu}_{test} = [2.7 \times 10^6 \textnormal{ g/(cm $\cdot$ s$^2$)}, 0.375]$ and  $t = 2.3084$ s. (\textbf{d}) $\boldsymbol{\mu}_{test} = [2.7 \times 10^6 \textnormal{ g/(cm $\cdot$ s$^2$)}, 0.375]$ and  $t = 2.64$ s.}
\label{fig7}
\end{figure}

In Figure~\ref{fig7} we report the FOM and the POD-DL-ROM velocity magnitudes, the latter with $N = 256$ and $n=5$, for two testing-parameter instances -- $\boldsymbol{\mu}_{test} = [2.3 \times 10^6 \textnormal{ g/(cm $\cdot$ s$^2$)}, 0.325]$ and $\boldsymbol{\mu}_{test} = [2.7 \times 10^6 \textnormal{ g/(cm $\cdot$ s$^2$)}, 0.375]$ -- at $t = 2.3084$ s and $t = 2.64$ s. 

\begin{figure}[h!]
\centering
\includegraphics[width=13 cm]{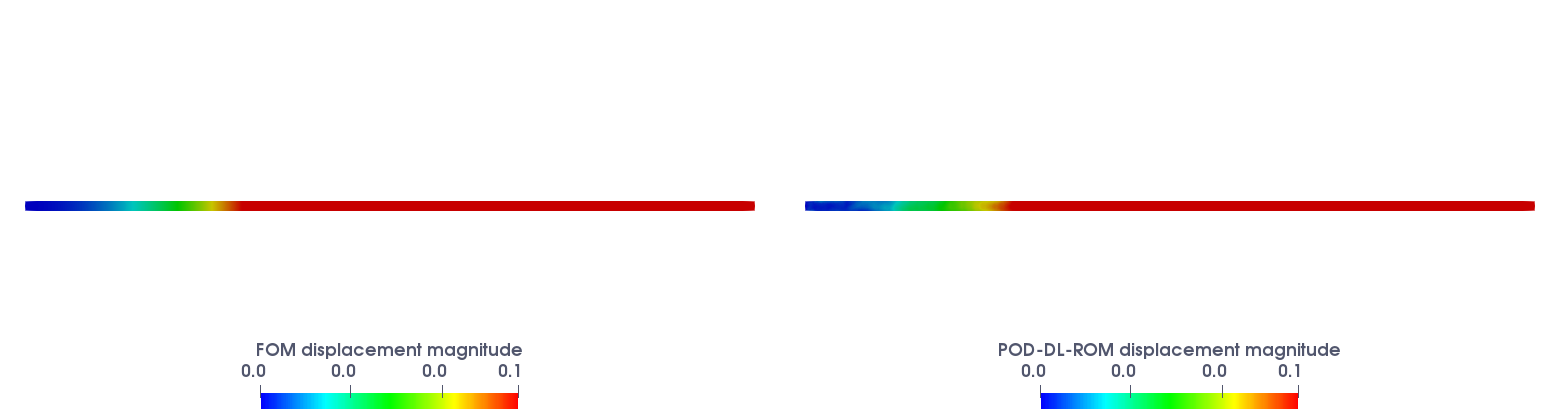}  \\
\includegraphics[width=13 cm]{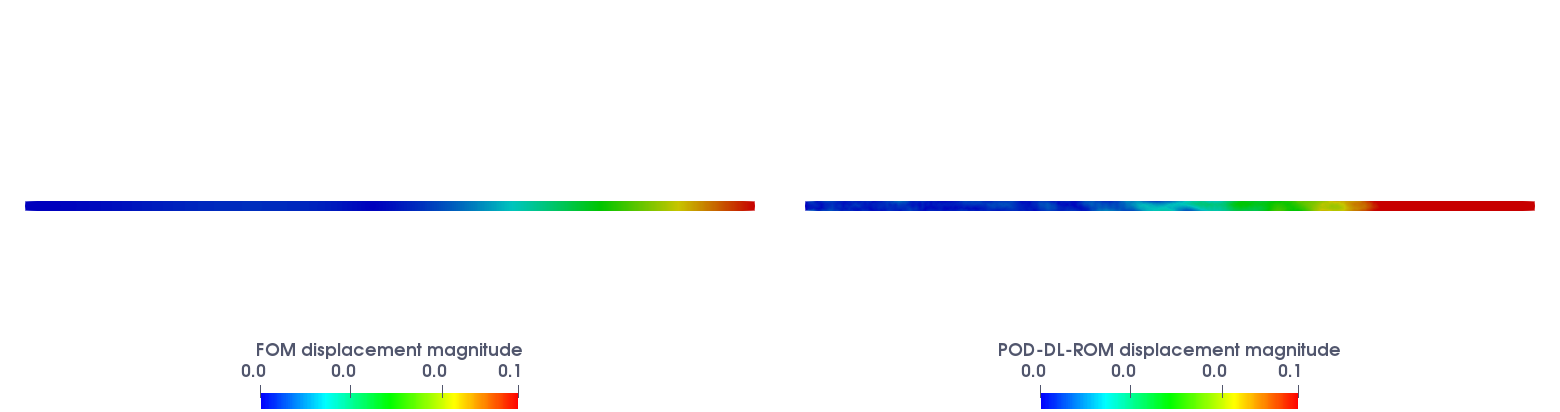} \\
\includegraphics[width=13 cm]{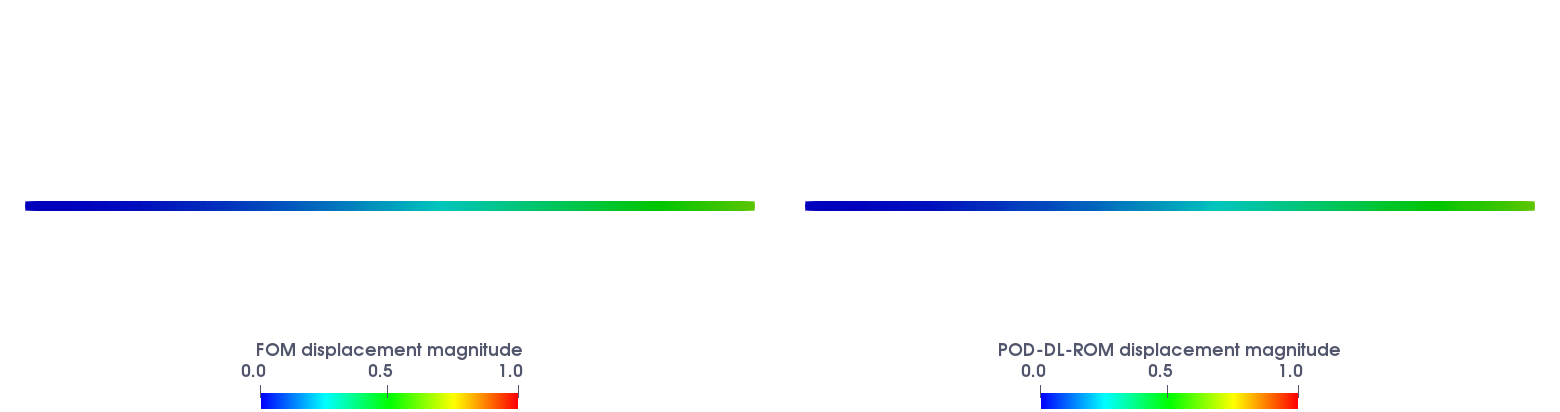} \\
\includegraphics[width=13 cm]{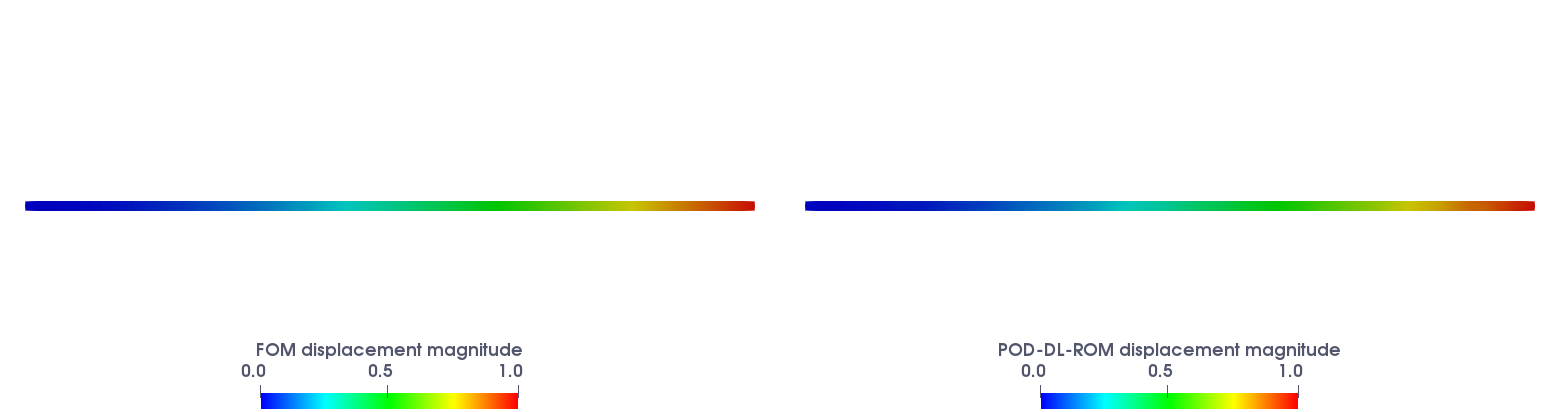}
\caption{\textit{Test case 2:} FOM and POD-DL-ROM structure displacement magnitudes 
 for the testing-parameter instances $\boldsymbol{\mu}_{test} = [2.3 \times 10^6 \textnormal{ g/(cm $\cdot$ s$^2$)}, 0.325]$ at $t = 2.3084$ s and $t = 2.64$ s (first and second row) and $\boldsymbol{\mu}_{test} = [2.7 \times 10^6 \textnormal{ g/(cm $\cdot$ s$^2$)}, 0.375]$ at $t = 2.3084$ s and $t = 2.64$ s (third and fourth row). }
 %for the testing-parameter instances $\boldsymbol{\mu}_{test} = [2.3 \times 10^6 \textnormal{ g/(cm $\cdot$ s$^2$)}, 0.325]$ and $\boldsymbol{\mu}_{test} = [2.7 \times 10^6 \textnormal{ g/(cm $\cdot$ s$^2$)}, 0.375]$ at $t = 2.3084$ s and $t = 2.64$ s, with $N = 256$ and $n = 5$. (\textbf{a}) $\boldsymbol{\mu}_{test} = [2.3 \times 10^6 \textnormal{ g/(cm $\cdot$ s$^2$)}, 0.325]$ and  $t = 2.3084$ s. (\textbf{b}) $\boldsymbol{\mu}_{test} = [2.3 \times 10^6 \textnormal{ g/(cm $\cdot$ s$^2$)}, 0.325]$ and  $t = 2.64$ s. (\textbf{c}) $\boldsymbol{\mu}_{test} = [2.7 \times 10^6 \textnormal{ g/(cm $\cdot$ s$^2$)}, 0.375]$ and  $t = 2.3084$ s. (\textbf{d}) $\boldsymbol{\mu}_{test} = [2.7 \times 10^6 \textnormal{ g/(cm $\cdot$ s$^2$)}, 0.375]$ and  $t = 2.64$ s.}
\label{fig8}
\end{figure}

We point out that the dependence of the displacement field on the parameters reflects on the velocity field by producing a strong variability over the parameter space, which is accurately captured by the POD-DL-ROM solutions. 
The FOM and POD-DL-ROM displacement magnitudes, for the testing-parameter instances $\boldsymbol{\mu}_{test} = [2.3 \times 10^6 \textnormal{ g/(cm $\cdot$ s$^2$)}, 0.325]$ and $\boldsymbol{\mu}_{test} = [2.7 \times 10^6 \textnormal{ g/(cm $\cdot$ s$^2$)}, 0.375]$ at $t = 2.3084$ s and $t = 2.64$ s over the domain, are shown in Figure~\ref{fig8}. The comparison between the FOM and POD-DL-ROM displacement magnitudes, for the testing-parameter instance $\boldsymbol{\mu}_{test} = [2.7 \times 10^6 \textnormal{ g/(cm $\cdot$ s$^2$)}, 0.375]$ at $\mathbf{x}^*=(5.50, 6.07)$ cm over time, is reported in Figure~\ref{fig8}, from which it is clearly visible that the POD-DL-ROM is also able to capture the main features of the elastic beam dynamics.

\begin{figure}[h!t]
\centerline{
\includegraphics[width=9.5 cm]{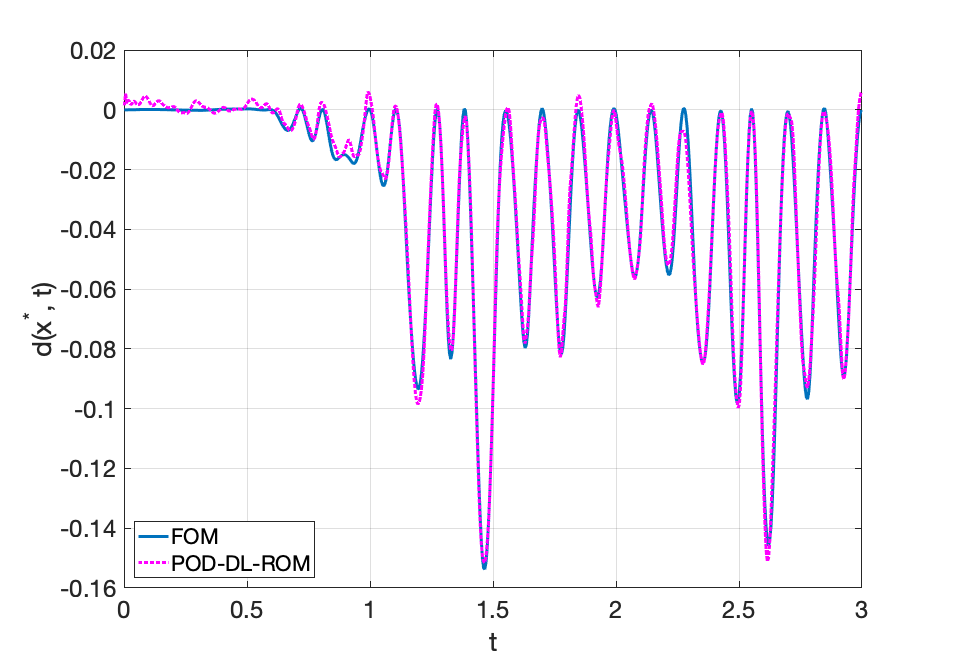} 
}
\caption{\textit{Test case 2:} FOM and POD-DL-ROM displacement at $\mathbf{x}^*=(5.50, 6.07)$ cm for the testing-parameter instance $\boldsymbol{\mu}_{test} = [2.7 \times 10^6 \textnormal{ g/(cm $\cdot$ s$^2$)}, 0.375]$.}
\label{fig9}
\end{figure}

Finally, in Table~\ref{tab1} we report the POD-DL-ROM GPU training (including validation) time, the testing time, i.e. the time needed to compute $N_t$ time instances for a testing-parameter instance, and the time required to compute one time instance at testing time. Indeed, we recall that the DL-ROM solution can be queried at a given time without requiring the solution of a dynamical system to recover the former time instances. We also show the speed-up gained by the POD-DL-ROM with respect to the CPU time needed to solve the FOM\footnote{For  test case 2, the FOM simulations have been carried out on a MacBook Pro Intel Core i7 6-core with 16 GB RAM CPU.}.

\begin{table}[h!] 
\caption{\textit{Test case 2:} POD-DL-ROM GPU computational times. \label{tab1}}
%%% \tablesize{} %% You can specify the fontsize here, e.g., \tablesize{\footnotesize}. If commented out \small will be used.
\centerline{
\begin{tabular}{cccc}
\hline
\textbf{training time [h]}	& \textbf{testing time [s]}	& \textbf{1-sample testing time [s]} & \textbf{speed-up}\\
\hline
7		& $4 \times 10^{-2}$			& $5 \times 10^{-3}$ & $1.77 \times 10^5$ ($1.41 \times 10^6$)\\
\hline
\end{tabular}
}
\end{table}

\subsection{Test case 3: blood flow in a cerebral aneurysm}

In this last test case we consider the fast simulation of blood flows in a cerebral (or intracranial) aneurysm, that is, a localized dilation or ballooning of a blood vessel in the brain, often occurring in the circle of Willis, the vessel network at the base of the brain. Blood velocity and pressure, wall shear stress (WSS),   blood flow impingement and particle residence time all play a key role in the growth and rupture of cerebral aneurysms -- see e.g. \cite{bazilevs_ane2010,cebral_2011,valencia_2008} -- which  might ultimately yield potentially severe brain damages. For these reasons, computational haemodynamics inside aneurysm models can provide output quantities of interest useful for planning their surgical treatment.

%Cerebral aneurysms are characterized by different sizes and shapes, and a broad distinction is made between saccular (the most common ones) and non-saccular aneurysms; the former are the most common ones and usually arise at a bifurcation or along a curve of the parent vessel. 

\begin{figure}[b!]
\centerline{
\includegraphics[width=6. cm]{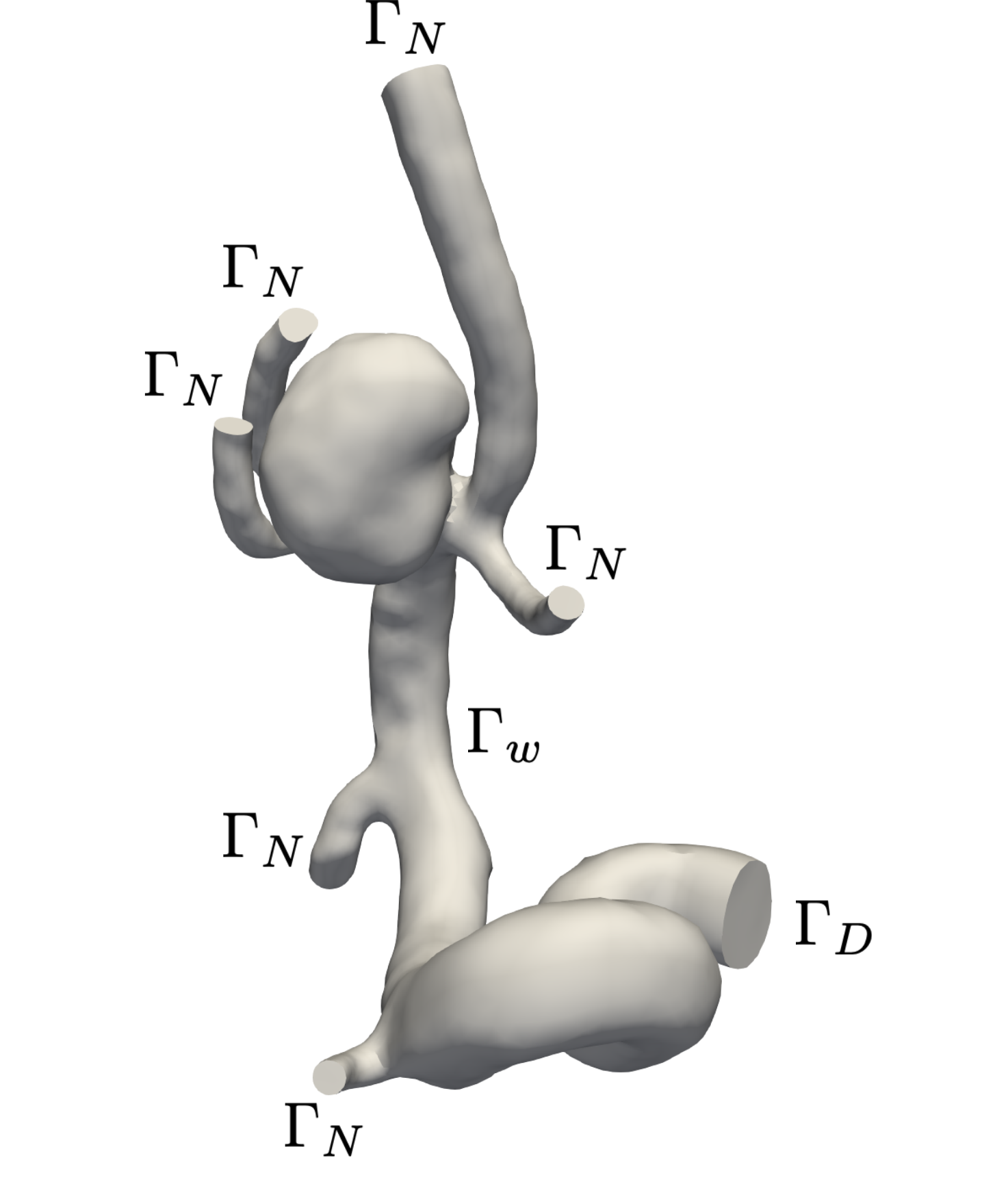} \hspace{-0.95cm}
\includegraphics[width=8.5 cm]{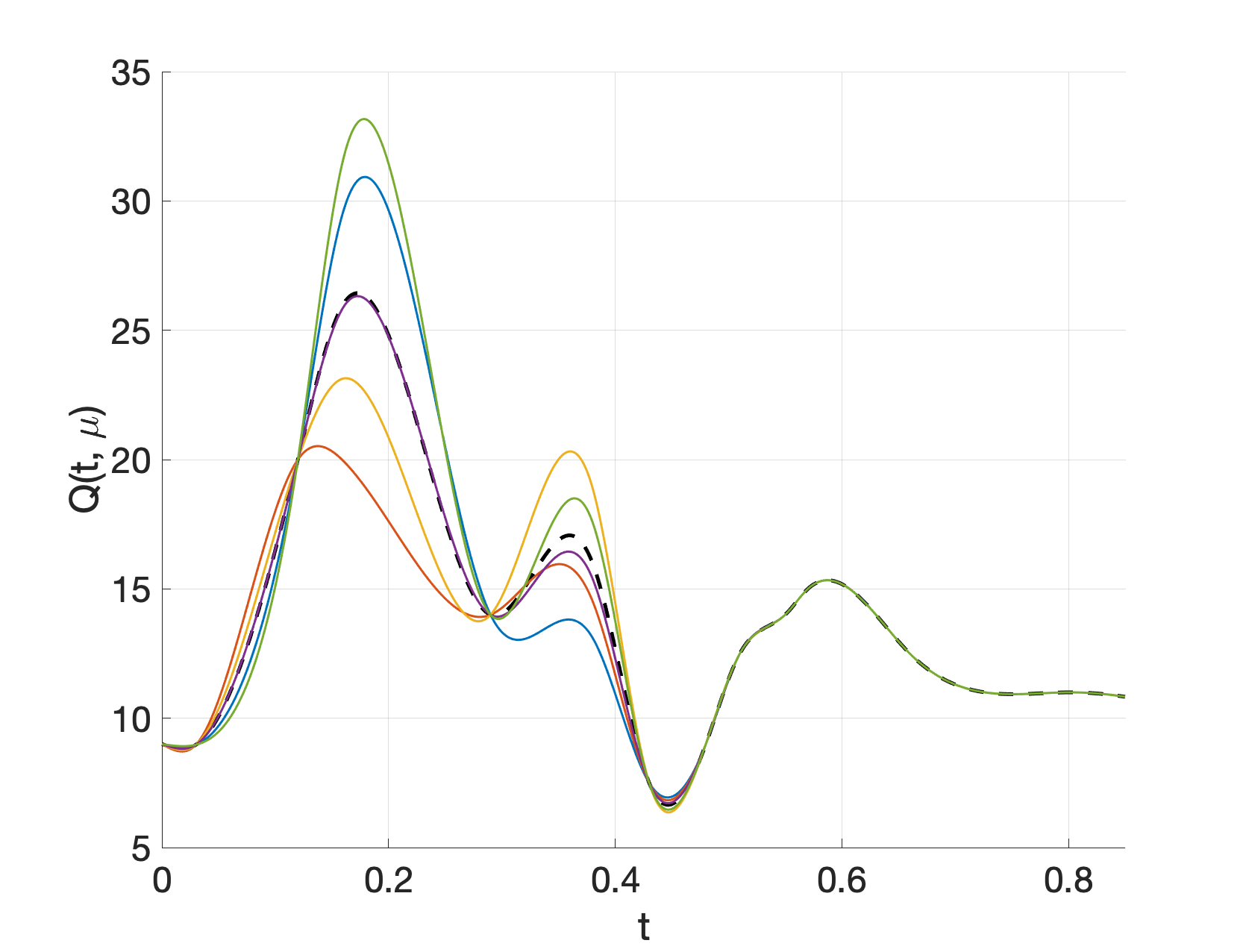} }
\caption{\textit{Test case 3. Left:} aneurysm geometry. 
\textit{Right:} inlet flow rate $Q(t; \boldsymbol{\mu})$ during the heart cycle for different parameter values; the black dashed curve corresponds to the base profile $Q(t)$.}
\label{fig9-0}
\end{figure}

We consider the artery aneurysm shown in Figure~\ref{fig9-0} (left), and extracted from the Aneurisk dataset repository \cite{aneurisk1,aneurisk2,piccinelli2009framework}. We consider blood as a Newtonian fluid, with constant viscosity, and a rigid arterial wall, so that blood flow dynamics can be described by the following Navier-Stokes equations:
\begin{equation}
\left\{
\begin{aligned}
& \rho \frac{\partial \mathbf{v}}{\partial t} + \rho \mathbf{v} \cdot \nabla \mathbf{v}   - \nabla \cdot \boldsymbol{\sigma}(\mathbf{v}, p) =  \mathbf{0} & \qquad  & (\mathbf{x},t) \in \Omega^F \times (0,T), \\
&\nabla \cdot \mathbf{v} = 0  & \qquad & (\mathbf{x},t) \in \Omega^F \times (0,T),\\
& \mathbf{v} =  \mathbf{0} & \qquad & (\mathbf{x},t) \in \Gamma_{w} \times (0,T),\\
& \mathbf{v} =  k\mathbf{v}_{in}Q(t) & \qquad & (\mathbf{x},t) \in \Gamma_{D} \times (0,T),\\
& \boldsymbol{\sigma}(\mathbf{v}, p) \mathbf{n}= \mathbf{0} & \qquad & (\mathbf{x},t) \in \Gamma_{N}\times (0,T), \\
&  \mathbf{v}(0) = \mathbf{0} & \qquad & \mathbf{x} \in \Omega, \ t = 0,
\end{aligned}
\right.
\label{eq:NS_aneurysm}
\end{equation}
where the stress tensor is defined as in (\ref{eq:sigma}). On the arterial wall $\Gamma_{w}$ a no-slip condition on the fluid velocity is imposed, flow resistance at the outlet boundaries $\Gamma_N$ is neglected, while a parabolic profile $\mathbf{v}_{in}$ is specified at the lumen inlet, where the parametrization of the inlet flow rate profile $Q(t; \boldsymbol{\mu})$ has been obtained by interpolating with radial basis functions a base profile $Q(t)$ taken from  \cite{blanco2015anatomical}, and then treating some of the interpolated values as parameters (see Figure~\ref{fig9-0}, right), see   \cite{negri2015efficient} for further details.

 In particular, we consider $n_{\mu} = 2$ parameters $\boldsymbol{\mu} \in \mathcal{P} \subset \mathbb{R}^2$ such that the flow rate at $t = 0.16$ s and $t = 0.38$ s admits variations up to 15$\%$ of the reference value. A comparison between some flow rate profiles corresponding to different parameter values is shown in Figure~\ref{fig9-0}, right. 
 %
%\begin{figure}[ht]
%\centering
%\includegraphics[width=9 cm]{aneurysm_params.png} \\
%\caption{\textit{Test 3:} Inlet flow rate $Q(t; \boldsymbol{\mu})$ during the heart cycle for different parameter values; the black dashed curve corresponds to the base profile $Q(t)$.}
%\label{fig10}
%\end{figure}
The scaling factor $k$ in (\ref{eq:NS_aneurysm}) is such that
\begin{equation*}
\int_{\Gamma_D} k \mathbf{v}_{in} \cdot \mathbf{n} d \sigma = 1.
\end{equation*}
Blood dynamic viscosity $\nu = 0.035$ P and density are set to $\rho = 1$ g/cm$^{3}$, respectively.

Concerning the FOM discretization, we employ a SUPG-BDF semi-implicit time
scheme of order 2 with linear finite elements for both velocity and pressure
variables. We employ a time-step $\Delta t = 10^{-3}$ over the interval $(0, T)$ with $T = 0.85$ s. We simulate the blood flow starting from an initial condition obtained by solving the steady Stokes problem. We are interested in reconstructing the blood velocity field, so we set the FOM dimensions to $N_h = [41985, 41985, 41985]$, and $p = 3$. The POD dimension is equal to $N = 64$, for each component of the solution, and the dimension of the reduced nonlinear trial manifold is chosen to be equal to $n = 5$, very close to the intrinsic dimension of the problem $n_{\mu} + 1 = 3$. We consider $N_t =850$ time instances, $N_{train} = 6$, and $N_{test} = 3$ training- and testing-parameter instances, sampled over $\mathcal{P}$ by means of  the latin hypercube sampling strategy.

\begin{figure}[h!]
\centering
\vspace{-0.25cm}
\includegraphics[width=15 cm]{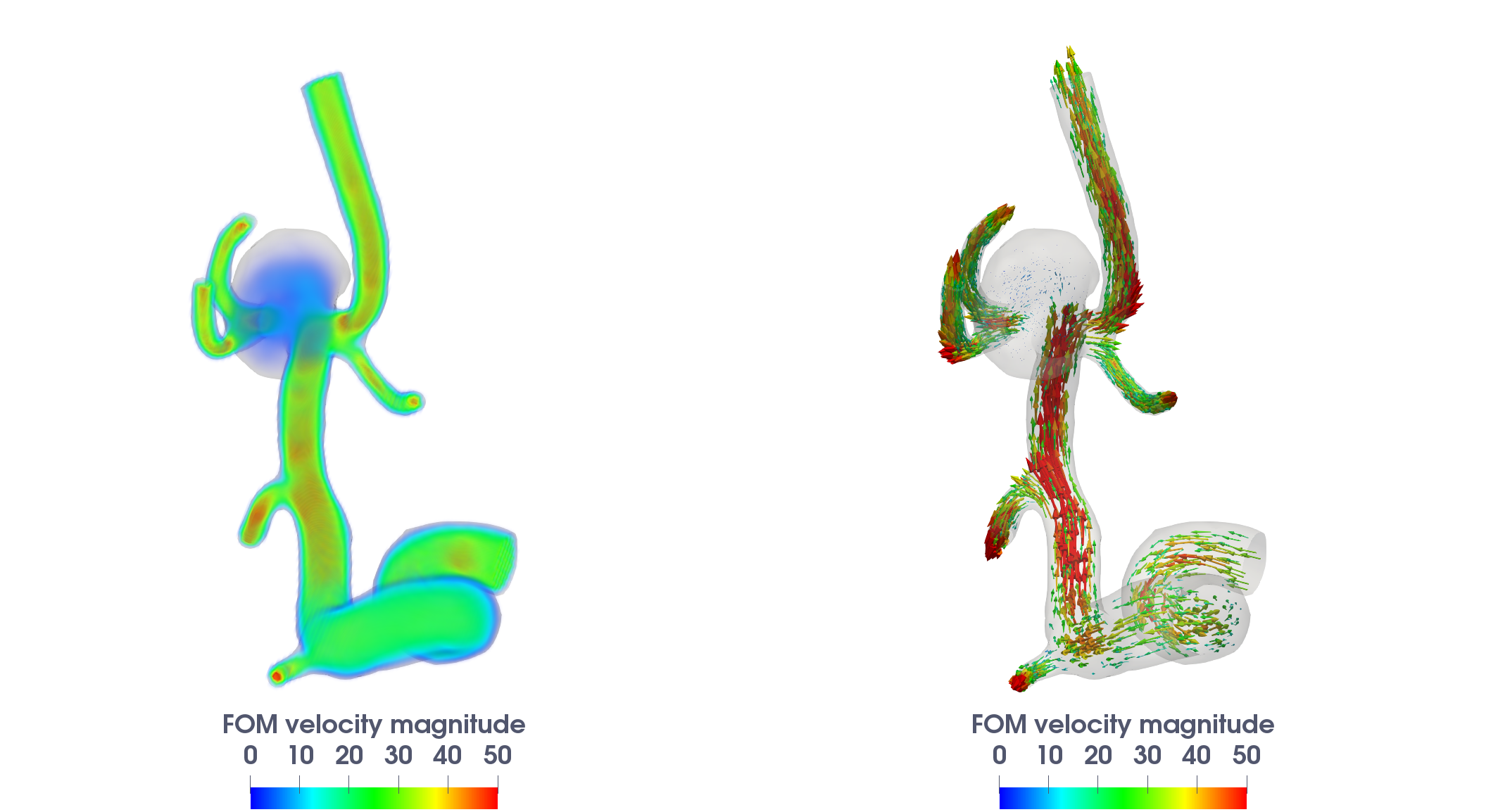}   \\
\includegraphics[width=15 cm]{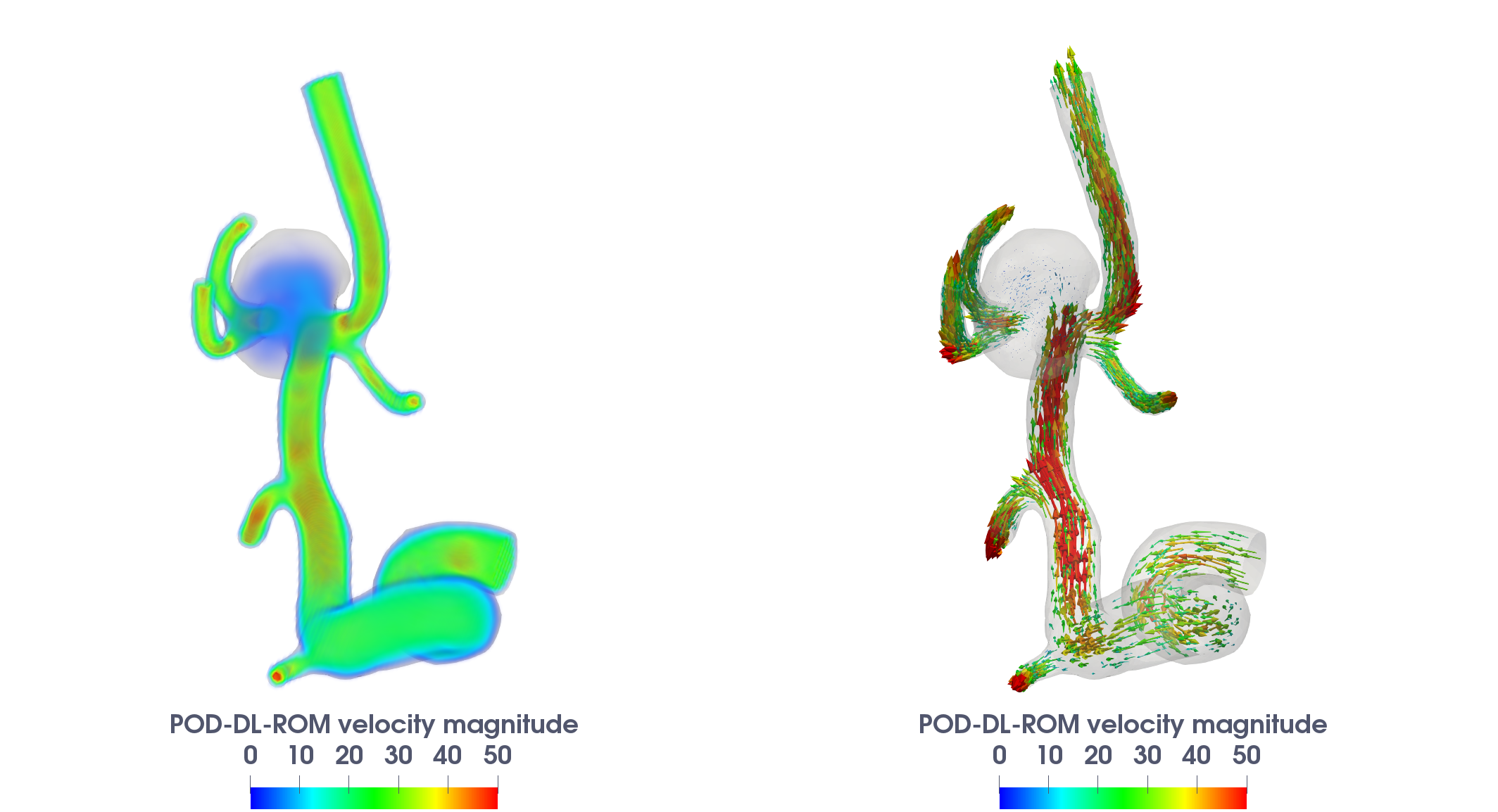} 
\caption{\textit{Test case 3:} FOM (top) and POD-DL-ROM (bottom) velocity fields  for the testing-parameter instance $\boldsymbol{\mu}_{test} = (5.9102, 3.1179)$ at the systolic peak $t = 0.18$ s.}
\label{fig11}
\end{figure}

\begin{figure}[h!t]
\centering
\vspace{-0.35cm}
\includegraphics[width=8.75 cm]{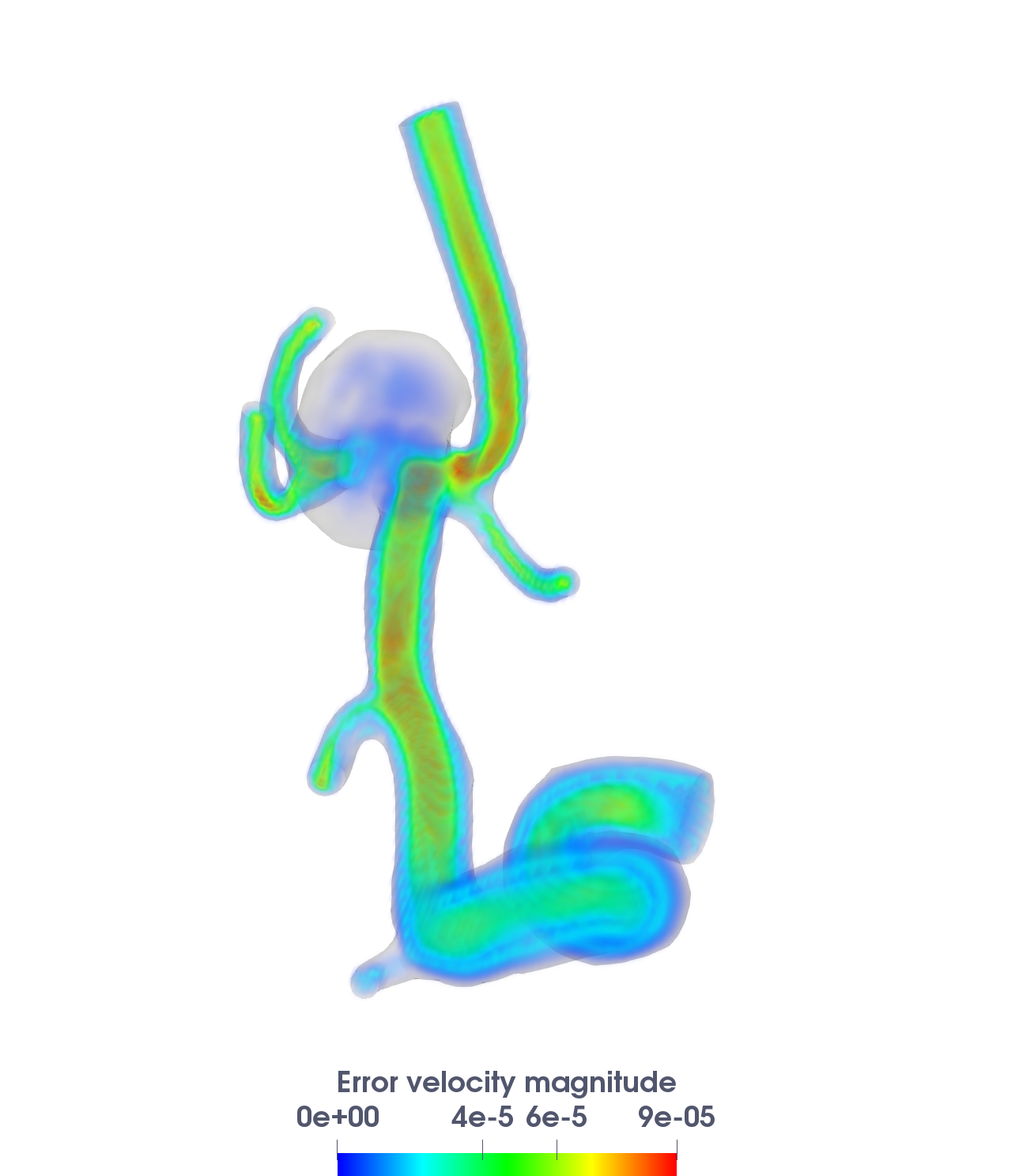}
\caption{\textit{Test case 3:} Relative error in the velocity magnitude for the testing-parameter instance $\boldsymbol{\mu}_{test} = (5.9102, 3.1179)$ at the systolic peak $t = 0.18$ s.}
\label{fig12}
\vspace{-0.25cm}
\end{figure}

In Figure~\ref{fig11} we compare the FOM and POD-DL-ROM velocity field magnitudes, the latter obtained with $N = 64$ and $n = 5$, for the testing-parameter instance $\boldsymbol{\mu}_{test} = (5.9102, 3.1179)$ at the systolic peak $t = 0.18$ s, along with the relative error $\boldsymbol{\epsilon}_k$ reported in Figure~\ref{fig12}. By looking at the pattern and the magnitude of the vector velocity field in Figure~\ref{fig11}, it is evident that the abnormal bulge and the inlet are the portions of the domain where the blood flow velocity is smaller, and we remark the ability of the POD-DL-ROM technique in capturing such dynamics in an extremely detailed manner.

\begin{figure}[h!t]
\centerline{
\includegraphics[width=12. cm]{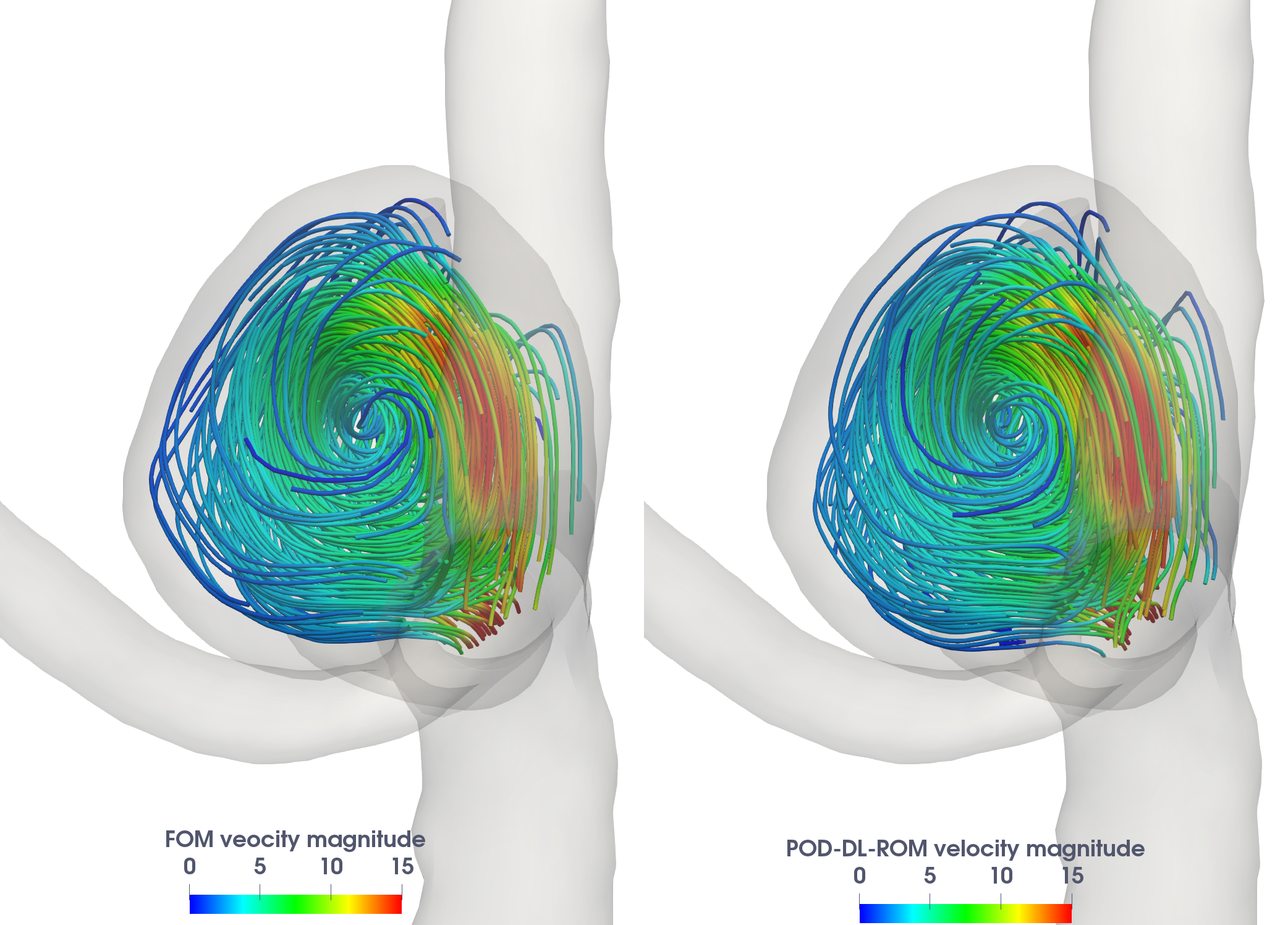}
}
%\vspace{0.25cm}
\caption{\textit{Test case 3:} FOM (left) and POD-DL-ROM (right) velocity magnitude streamlines for the testing-parameter instance $\boldsymbol{\mu}_{test} = (5.9102, 3.1179)$ at the sistolic peak $t = 0.5$ s. }
\label{fig13}
\end{figure}

\begin{figure}[h!]
\centering
\includegraphics[width=12.5 cm]{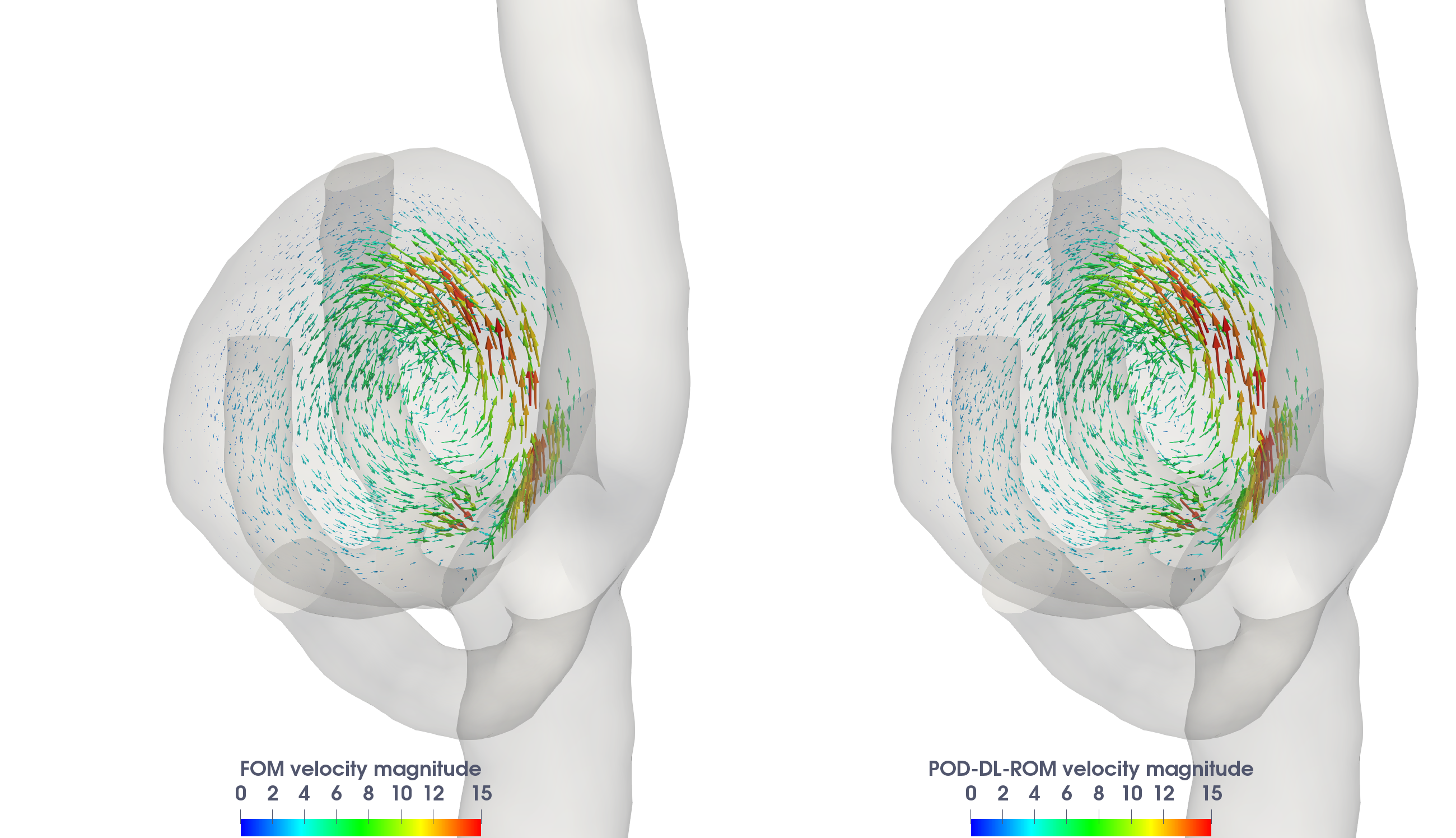} \\
\caption{\textit{Test case 3:} FOM (left) and POD-DL-ROM (right) velocity field magnitude for the testing-parameter instance $\boldsymbol{\mu}_{test} = (5.9102, 3.1179)$ at  $t = 0.18$ s.}
\label{fig14}
\end{figure}

In Figure~\ref{fig13} we report the streamlines of the blood velocity field, obtained with the  FOM and the POD-DL-ROM, for the testing-parameter instance $\boldsymbol{\mu}_{test} = (5.9102, 3.1179)$ at  $t = 0.5$ s. In Figure~\ref{fig14} we report instead a detailed view of the pattern of the fluid velocity field obtained  for the testing-parameter instance $\boldsymbol{\mu}_{test} = (5.9102, 3.1179)$ at $t = 0.18$ s,  highlighting the recirculation of the flow in the bulge and the blood stasis in this region.

Finally, the testing computational time, i.e. the time needed to compute $N_t$ time instances for an unseen testing-parameter instance, of the POD-DL-ROM on a Tesla V100 32GB GPU is given by 0.28 seconds, thus implying a speed-up equal to $3.98 \times 10^5$ with respect to the time needed for the solution of the FOM\footnote{For test case 3, the  FOM simulations have been carried out on 20 cores of 1.7 TB node (192 Intel\textsuperscript{\textregistered} Xeon Platinum\textsuperscript{\textregistered} 8160 2.1GHz cores) of the HPC cluster available at MOX, Politecnico di Milano.}, and the possibility to obtain a fully detailed simulation of a complex blood flow in real-time.

%%%%%%%%%%%%%%%%%%%%%%%%%%%%%%%%%%%%%%%%%%

\section{Discussion}\label{sec:discussion}

In this work, we have taken advantage of a recently proposed technique \cite{fresca2020POD} to build non-intrusive low-dimensional ROMs by exploiting DL algorithms to handle fluid dynamics problems. This strategy allows us to overcome some drawbacks of classical projection-based ROM techniques arising when they are applied to incompressible flow simulations. 

In particular, POD-DL-ROMs overcome the need of:
\begin{itemize}
\item treating efficiently nonlinearities and (nonaffine) parameter dependencies, thus avoiding expensive and intrusive hyper-reduction techniques; 
\item approximating both velocity and pressure fields, in those cases where one might be interested only in the visualization of a single field;
\item imposing physical constraints that couple different submodels, as in the case of fluid-structure interaction (the different field variables are indeed treated as independent by the neural network);
\item ensuring the ROM stability by enriching the reduced basis spaces;  
\item solving a dynamical system at the reduced level to model the fluid dynamics, however keeping the error propagation in time under control.
\end{itemize}

We assessed the performance of the POD DL-ROM technique on three test cases, dealing with the flow around a cylinder benchmark, the fluid-structure interaction between an elastic beam attached to a fixed, rigid block and a laminar incompressible flow, and the blood flow in a cerebral aneurysm,  by showing its ability in providing accurate and efficient (even on moderately  large-scale problems) ROMs, which multi-query and real-time applications may ultimately rely on. In particular, the prior dimensionality reduction performed through POD on the snapshot matrices also enhances the overall efficiency of the technique during the offline training stage. 

Therefore, we can conclude that POD-DL-ROMs provide a non-intrusive and general-purpose tool enabling us to perform real-time  numerical simulations of fluid flows. Since they return a (FOM-like detailed) computation of the field variables, rather than approximating selected output quantities of interest as in the case of traditional emulators or surrogate models, POD-DL-ROMs are a viable tool for detailed flow analysis, without any requirement in terms of computational resources during the online testing stage. 

%Authors should discuss the results and how they can be interpreted from the perspective of previous studies and of the working hypotheses. The findings and their implications should be discussed in the broadest context possible. Future research directions may also be highlighted.

%%%%%%%%%%%%%%%%%%%%%%%%%%%%%%%%%%%%%%%%%%
\vspace{6pt} 

%%%%%%%%%%%%%%%%%%%%%%%%%%%%%%%%%%%%%%%%%%
%% optional
%\supplementary{The following are available online at \linksupplementary{s1}, Figure S1: title, Table S1: title, Video S1: title.}

% Only for the journal Methods and Protocols:
% If you wish to submit a video article, please do so with any other supplementary material.
% \supplementary{The following are available at \linksupplementary{s1}, Figure S1: title, Table S1: title, Video S1: title. A supporting video article is available at doi: link.} 

%%%%%%%%%%%%%%%%%%%%%%%%%%%%%%%%%%%%%%%%%%
\section*{Authors' Contributions}
Conceptualization: S.F. and A.M.;  Formal analysis: S.F.;  Funding acquisition: A.M.; Investigation: S.F.;  Methodology: S.F. and A.M.; Software: S.F.; Supervision: A.M.; Validation: S.F.;  Visualization: S.F.; Writing: S.F. and A.M. .

\section*{Acknowledgements} 
This work has been supported by Fondazione Cariplo (grant agreement no. 2019 - 4608, P.I. Prof. A. Manzoni).

\bibliography{references}

% If authors have biography, please use the format below
%\section*{Short Biography of Authors}
%\bio
%{\raisebox{-0.35cm}{\includegraphics[width=3.5cm,height=5.3cm,clip,keepaspectratio]{Definitions/author1.pdf}}}
%{\textbf{Firstname Lastname} Biography of first author}
%
%\bio
%{\raisebox{-0.35cm}{\includegraphics[width=3.5cm,height=5.3cm,clip,keepaspectratio]{Definitions/author2.jpg}}}
%{\textbf{Firstname Lastname} Biography of second author}

% The following MDPI journals use author-date citation: Arts, Econometrics, Economies, Genealogy, Humanities, IJFS, JRFM, Laws, Religions, Risks, Social Sciences. For those journals, please follow the formatting guidelines on http://www.mdpi.com/authors/references
% To cite two works by the same author: \citeauthor{ref-journal-1a} (\citeyear{ref-journal-1a}, \citeyear{ref-journal-1b}). This produces: Whittaker (1967, 1975)
% To cite two works by the same author with specific pages: \citeauthor{ref-journal-3a} (\citeyear{ref-journal-3a}, p. 328; \citeyear{ref-journal-3b}, p.475). This produces: Wong (1999, p. 328; 2000, p. 475)

%%%%%%%%%%%%%%%%%%%%%%%%%%%%%%%%%%%%%%%%%%
%% for journal Sci
%\reviewreports{\\
%Reviewer 1 comments and authors’ response\\
%Reviewer 2 comments and authors’ response\\
%Reviewer 3 comments and authors’ response
%}
%%%%%%%%%%%%%%%%%%%%%%%%%%%%%%%%%%%%%%%%%%
\end{document}